\documentclass[1p,preprint,10pt]{elsarticle}
\usepackage{amsthm}
\usepackage{amssymb}
\usepackage{mathtools}
\usepackage[T1]{fontenc} 
\usepackage{amsmath} 
\newtheorem{theorem}{Theorem}[section]

\newtheorem{remark}{Remark}[section]
\newtheorem{lemma}{Lemma}[section]

\theoremstyle{definition}
\newtheorem{definition}{Definition}[section]
\usepackage{subfigure}
\usepackage{hyperref}
\usepackage[numbers]{natbib}
\biboptions{sort&compress}
\usepackage{booktabs}
\usepackage[utf8]{inputenc} 
\newcommand{\transp}{\text{T}}
\usepackage{xcolor}

\begin{document}

\begin{frontmatter}

\title{A boostlet transform for wave-based acoustic signal processing in space--time}

\author[1]{Elias Zea\corref{cor1}}
\cortext[cor1]{Corresponding author}
\ead{zea@kth.se}
\address[1]{KTH Royal Institute of Technology, Department of Engineering Mechanics, The Marcus Wallenberg Laboratory for Sound and Vibration Research, Teknikringen 8, SE-114 28 Stockholm, Sweden}

\author[2]{Marco Laudato}
\address[2]{KTH Royal Institute of Technology, Department of Engineering Mechanics, FLOW Centre, Osquars Backe 18, SE-114 28 Stockholm, Sweden}

\author[3]{Joakim And{\'e}n}
\address[3]{KTH Royal Institute of Technology, Department of Mathematics, Division of Probability, Mathematical Physics and Statistics, Lindstedtvägen 25, SE-114 28 Stockholm, Sweden}

\begin{abstract}
Sparse representation systems that encode signal architecture have had a profound impact on sampling and compression paradigms. Remarkable examples are multi-scale directional systems, which, similar to our vision system, encode the underlying architecture of natural images with sparse features. Inspired by this philosophy, we introduce a representation system for wave-based acoustic signal processing in 2D space--time, referred to as the \emph{boostlet transform}, which encodes sparse features of natural acoustic fields using the Poincaré group and isotropic dilations. Boostlets are spatiotemporal functions parametrized with dilations, Lorentz boosts, and translations in space--time. Physically speaking, boostlets are supported away from the acoustic radiation cone, i.e., having broadband frequency with phase velocities other than the speed of sound, resulting in a peculiar scaling function. We formulate a discrete boostlet frame using Meyer wavelets and bump functions and examine its sparsity properties. An analysis with experimentally measured fields indicates that discrete boostlet coefficients decay significantly faster and attain superior reconstruction performance than wavelets, curvelets, shearlets, and wave atoms. The results demonstrate that boostlets provide a natural, compact representation system for acoustic waves in space-time.  
\end{abstract}


\end{frontmatter}

\section{Introduction}

Wave equations are mathematical structures central in many fields of physics and engineering, with applications ranging from medical imaging to acoustics, seismic exploration, electromagnetism, and gravitational wave detection. In acoustics, canonical solutions to the wave equation are well-known in various geometries~\cite{Morse1968}, typically described in planar, cylindrical, and spherical coordinates. From Fourier's uncertainty principle, planar wave solutions are fully localized in the wavenumber-frequency domain and non-localized in space--time. As a consequence, modeling localized phenomena in space--time---such as wave scattering of an object with a size comparable to the wavelength---is challenging and computationally expensive. A large number of expansion coefficients are needed to model the spatiotemporal wave decay in the near-field. Another challenging aspect of acoustic scattering from a finite object is the transition of the wave from being localized in space--time in the vicinity of the object, with broadband Fourier spectrum, to being spatially extended far away from the object, with band-limited Fourier spectrum. As a consequence of the Sommerfeld radiation boundary condition~\cite{Sommerfeld1912}, the compact wave expands spherically outwards and turns into a transient planar wave---becoming a localized ridge in the Fourier domain. How these dynamics are captured by a representation system is key for analyzing and processing acoustic fields in space--time. 

\subsection{From Gabor atoms to Poincaré wavelets}

Decades of efforts in applied harmonic analysis have led to remarkable representation systems. In 1946, Gabor presented a theory for time--frequency analysis~\cite{Gabor1946} by applying Heisenberg's uncertainty principle and establishing an analogy between sound and quanta. Gabor envisioned that a family of parametrized \emph{atoms} could detect transient components more effectively than Fourier bases. Gabor's ideas later triggered the development of the wavelet transform in the 1970s, coined initially as the cochlear transform by Zweig~\cite{Zweig1976}, and later formalized in various ways by Grossmann, Morlet, Debauchies, and Meyer in the 1980s--1990s~\cite{Grossmann1984, Daubechies1992, Meyer1993}. 

Notable works in the early 1990s by Ali, Antoine, and Gazeau established a framework for square-integrability conditions of unitary representations in space--time. In a two-part study, they introduced reproducing triples and continuous frames~\cite{Ali1991a} and constructed explicit coherent-state wavelet bases on the forward light-cone~\cite{Ali1991b}, laying the mathematical foundation for analyzing wave equations via continuous wavelet transforms. Related results were found by Klauer and Streater on cyclic representations of the Poincaré group with a resolution of identity~\cite{Klauder1991}, and by Bohnke on the admissibility conditions for Lorentz-group wavelets and tight frames of localized solutions on the light cone~\cite{Bohnke1991}. Bernier and Taylor later generalized these ideas for square integrable representations of semidirect products of groups~\cite{Bernier1996}. These works were largely theoretical, but provided a basis for later applications and were eventually consolidated by Ali, Antoine, and Gazeau in a comprehensive book on coherent states and wavelets~\cite{Ali2000}.

By the mid-1990s, Kaiser had introduced the notion of physical wavelets as actual acoustic or electromagnetic pulses rather than abstract basis functions~\cite{Kaiser2003,Kaiser2011}. Kiselev, Perel, and Sidorenko built on these foundations by formulating the Poincaré wavelet transform~\cite{Kiselev2000Localized,Perel2009Integral}, comprising wavelets that are dilated, Lorentz-boosted, and translated in space--time. In the field of image processing, many other anisotropic transforms were constructed in the late 1990s and early 2000s, e.g., directional wavelets to localize edges~\cite{Antoine1996}, ridgelets to localize ridge functions~\cite{Candes1999a}, and curvelets~\cite{Candes2004}, shearlets~\cite{Labate2005}, and contourlets~\cite{Do2005} to localize curved singularities. In the mid 2000s, research in microlocal analysis showed that curvelets and shearlets resolve the wavefront set~\cite{Candes2005a, Kutyniok2008}. In fact, curvelets~\cite{Candes2005}, shearlets~\cite{Guo2008}, and wave atoms~\cite{Demanet2009} provide optimally sparse representations of the wave propagator in free space. In the late 2000s, Demanet and Ying~\cite{Demanet2009} showed that optimally sparse representations of Fourier integral operators and Green's function of the wave equation are attained with anisotropic essential support of the order of $2^{-aj}\times 2^{-j/2}$, where $a \in [1/2,1]$. At around the same time, Pinto and Vetterli introduced directional filter banks to approximate wavefronts measured by smoothly-varying 1D microphone arrays~\cite{Pinto2010,Pinto2011}.

\subsection{Historical developments in acoustic simulation and experiments}

The analysis and processing of acoustic fields has seen numerous developments since the first computer simulations in the late 1950s~\cite{Allred1958, Schroeder1962}, when applications of geometrical-acoustics methods, e.g., ray tracing~\cite{Krokstad1968}, quickly became popular. Some years later, the image source method was developed as an analytical tool to predict acoustical reflections in complex room geometries~\cite{Allen1979}. In the 1980s, Williams and collaborators made monumental advancements in sound source quantification by introducing acoustic holography~\cite{Williams1980}, an industrial breakthrough for contactless characterization of vibroacoustic sources. By the end of the 1990s, Berkhout \emph{et al.} studied the analysis~\cite{Berkhout1997} and extrapolation~\cite{Berkhout1999} of acoustic fields inside rooms in great detail by using analogies from seismic exploration. Numerical approaches such as finite elements~\cite{Craggs1994}, finite-difference-time-domain methods~\cite{Botteldooren1995}, and, more recently, spectral element methods~\cite{Pind2019} and discontinuous Galerkin methods~\cite{Wang2019}, have since been used extensively to approximate the solution by discretizing the computational domain. 

Since the emergence of compressed sensing, sparse representations have had impactful applications in acoustic simulations and experiments, including super-resolution, multi-scale analysis, and feature extraction for model-based and machine-learning frameworks. Sampling and reconstructing acoustic fields in space--time is challenging because many data points are required to avoid aliasing. For example, in the human audible range ($20$~Hz to $20$~kHz), about $1.6$~million microphone positions are necessary to reconstruct the pressure field over a cubic meter of air. Thus, overcoming sampling limitations has a profound impact on technologies based on the wave equation, such as seismic exploration, medical ultrasound, and vibroacoustic imaging, to mention a few. Notable studies have investigated the sparse reconstruction of acoustic fields with monopole sources~\cite{Mignot2013, Antonello2017}, plane waves~\cite{Verburg2018}, curvelets~\cite{Herrmann2008}, and shearlets~\cite{Zea2019}. Finally, data-driven and machine-learning methods, such as physics-informed neural networks~\cite{Shukla2020, Borrel-Jensen2021, Rasht-Behesht2022, Wang2023,Karakonstantis2024}, deep prior approaches~\cite{Siahkoohi2020, Kong2022, Pezzoli2022}, generative models~\cite{Dietrichson2018, Goudarzi2020, Fernandez-Grande2023, Karakonstantis2023a}, and neural operators~\cite{Dai2023, Middleton2023, Sun2023, Borrel-Jensen2024} have recently achieved impressive results due to their ability to more closely adapt to the structure (spatial or temporal) of the acoustic fields. 

\subsection{Contribution of the work}

It is worth noting how this study is placed in the existing literature. The mathematical structure of the boostlet transform is similar to wavelet constructions based on the Poincaré group~\cite{Ali2000}. In particular, boostlets are spatiotemporal atoms parameterized by continuous dilations, Lorentz boosts, and space--time translations---mirroring the parameters of the Poincaré wavelet transform. Despite this similarity, Poincaré-based systems were originally studied in the context of relativistic monochromatic point sources~\cite{Perel2012Relativistic} and asymptotic solutions to the wave equation in smoothly inhomogeneous media~\cite{Gorodnitskiy2023}. In the former context, a Lorentz boost $\theta$ is interpreted from a relativistic perspective as an observer's velocity, e.g., as in optics, which is inapplicable to acoustic, wave-like signals. In the latter context, a Lorentz boost is interpreted as a characteristic ray that is traced in the medium via $\tan{\beta} = \sinh{\theta}$, which is an asymptotic assumption valid for particle-like (high-frequency) wave propagation. In contrast, the boostlet transform proposed in this work decomposes broadband wavefronts in space--time directly in terms of their phase speed content---determined by passbands of Lorentz boosts---and their frequency content---determined by passbands of space--time scales. Additionally, the boostlet transform takes into consideration finite scales, completing a resolution of identity and thereby ensuring perfect broadband wavefield reconstruction down to zero hertz.

\begin{figure}[!ht]
    \centering
    \includegraphics[width=0.95\linewidth]{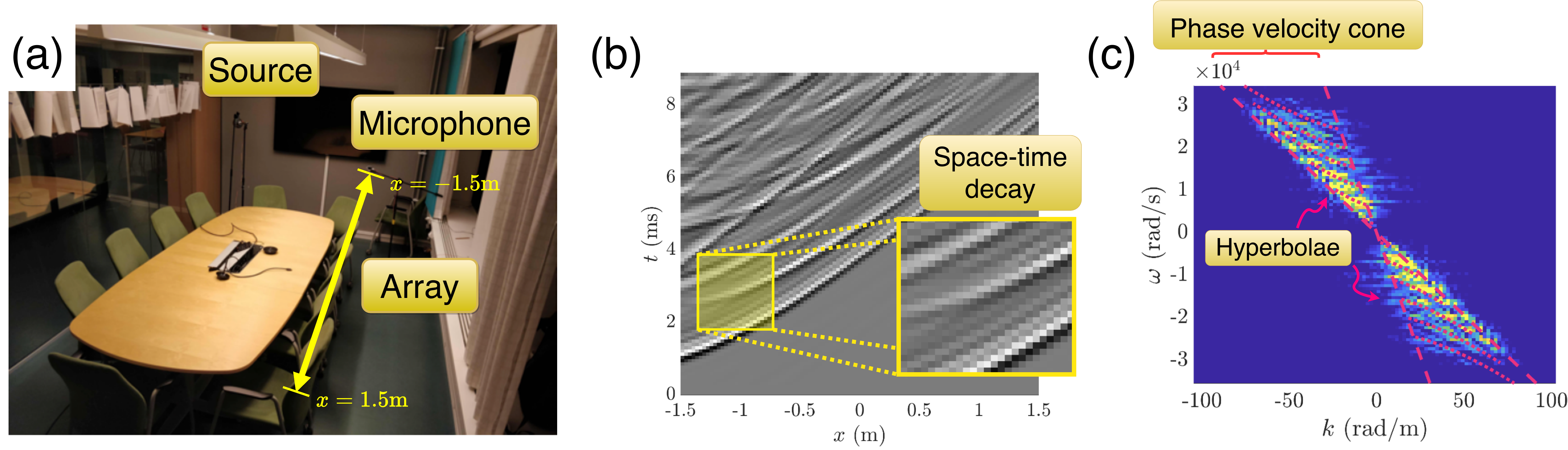}
    \caption{Illustration of a room-acoustical measurement in 2D space--time. (a) Experimental setup with a sound source emitting a frequency sweep between $100$~Hz and $4.5$~kHz and a free-field microphone recording the sound pressure. (b) Sound pressure field measured for approximately $8.9$~ms on a line of $100$~microphone positions spanning $3$~m. The zoom-in shows a wave decaying in space--time, entailing a scattered wave. (c) Fourier spectrum of (b). The spectrum shows several hyperbolae localized in a phase velocity cone (dashed lines shown for illustration purposes).}
    \label{fig:setup}
\end{figure}

The main contributions of this work are summarized in what follows:
\begin{itemize}
    \item[(i)] We apply the principle of analogy from sparsely coding natural images~\cite{Olshausen1996, Donoho2001} to sparsely coding acoustic waves in space--time. Previous theoretical and experimental results~\cite{Zea2021} indicate that natural acoustic fields in 2D space--time can be sparsely encoded into band-limited wave shapes supported on frequency-phase speed passbands. This is shown in Figure~\ref{fig:setup}(c), where anti-symmetric hyperbolae localized in phase velocity cones appear to characterize these shapes. This provides experimental evidence supporting that the Poincaré group is a suitable candidate to sparsely represent broadband wavefields in space--time. In particular, this contribution extends to feature-based machine learning in the sense that boostlet features no longer need to be learned during training.
    
    \item[(ii)] We introduce the \emph{boostlet transform}, which decomposes broadband acoustic waves into a collection of band-limited functions on $L^2(\mathbb{R}^2)$, referred to as \emph{boostlets}, parametrized by the Poincar{\'e} group (Lorentz boosts and translations), and isotropic dilations in 2D space--time. The physical meaning of the admissibility condition and the associated scaling function are discussed. A discrete formulation with a Meyer wavelet system is presented. When applied to acoustic fields measured experimentally, discrete boostlet decompositions have a faster coefficient decay compared to benchmark systems such as wavelets, shearlets, and wave atoms.
\end{itemize}

In addition to the abovementioned contributions, the present work aims to improve the accessibility and applicability of space--time wavelets in the context of engineering sciences. While the mathematical foundation of wavelets built from unitary representations of the Poincaré group has been studied since the early 1990s~\cite{Ali1991a,Ali1991b,Bohnke1991,Ali2000,Kaiser2003,Kiselev2000Localized,Perel2009Integral,Gorodnitskiy2011}, much of this literature is practically inaccessible for researchers working outside harmonic analysis and mathematical physics. In this way, the present work contributes to making space--time wavelet systems more usable and interpretable within the engineering sciences.

\section{Background}

Although solutions of the wave equations have a wide frequency range, this study focuses on the acoustic regime between infrasound and ultrasound. Our ambition is that the methodology can be extended to other types of waves and propagation media. This section presents the relevant background on canonical solutions to time-harmonic wave equations, Fourier acoustics theory, encoding dispersion relations with the Poincaré group, and discrete frame theory preliminaries. 

\subsection{Wave equation preliminaries}
\label{sec:waveeqs}
The linearized wave equation in a homogeneous medium with constant sound speed, in the absence of flow, is a second-order hyperbolic partial differential equation (PDE) of the form
\begin{equation}
\label{eq:waveeq}
    \Box p(\mathbf{r},t) : = \left( \nabla^2 - \frac{1}{c_0^2} \frac{\partial^2}{\partial t^2}\right) p(\mathbf{r},t) = 0,
\end{equation}
where $\Box$ is known as the d'Alembert operator, $p(\mathbf{r},t) \in \mathbb{R}$ is the acoustic pressure at the position $\mathbf{r}$ and time $t$, $c_0 \in \mathbb{R}^+$ is the sound speed of the medium (e.g., air at room temperature), and $\nabla^2$ is the Laplacian operator, defined here in Cartesian coordinates as
\begin{equation}
    \nabla^2 = \frac{\partial^2}{\partial x^2} + \frac{\partial^2}{\partial y^2} + \frac{\partial^2}{\partial z^2}.
\end{equation}

A classical approach to solving the wave equation is to separate the space and time variables and compute the stationary wave solutions by fixing the angular frequency $\omega = \omega_0 \in \mathbb{R}^+$, giving
\begin{equation}
    p(\mathbf{r},t) = \hat{p} (\mathbf{r};\omega_0) e^{-i \omega_0 t},
\end{equation}
where $i$ is the imaginary unit. The separation of variables turns the hyperbolic PDE in Eq.~\eqref{eq:waveeq} into the elliptic PDE 
\begin{equation}
\label{eq:helmholtz}
    \nabla^2 \hat{p}(\mathbf{r};\omega_0) + k^2 \, \hat{p}(\mathbf{r};\omega_0) = 0,
\end{equation}
commonly known as the Helmholtz equation in acoustics and electromagnetism. Here, the acoustic wavenumber is $k_0 = \omega_0/c_0$; i.e., acoustic waves are linearly dispersive. A common solution in Cartesian coordinates for Eq.~\eqref{eq:helmholtz} is a plane wave solution of the form
\begin{equation}
\label{eq:ansatz}
\hat{p}(\mathbf{r};\omega_0) = P(\omega_0) \, e^{i \, (\mathbf{r} \cdot \hat{\mathbf{k}}) }, 
\end{equation}
where $P(\omega_0) \in \mathbb{C}$ is the amplitude of the wave at frequency $\omega_0$, $(\, \cdot \,)$ denotes the scalar product, and $\hat{\mathbf{k}} = (\hat{k}_x,\hat{k}_y,\hat{k}_z)$ is the wavenumber vector with components defined as:
\begin{equation}
\label{eq:kxkykz}
    \begin{split}
        \hat{k}_x & = k_0 \cos\alpha \sin\beta, \\
        \hat{k}_y & = k_0 \sin\alpha \cos\beta, \\
        \hat{k}_z & = k_0 \cos\beta,
    \end{split}
\end{equation}
where $(\alpha,\beta)$ are the azimuth and elevation angles, respectively, along which the wave propagates.

The spatial Fourier transform of Eq.~\eqref{eq:ansatz} gives the following Dirac delta located at $(\hat{k}_x,\hat{k}_y,\hat{k}_z)$
\begin{equation}
    \hat{p}(\mathbf{k};\omega_0) = \int_{\mathbb{R}^3} P(\omega_0) e^{i \, (\mathbf{r} \cdot \hat{\mathbf{k}})} e^{-i(\mathbf{r} \cdot \mathbf{k})} \mathrm{d}\mathbf{r} = P(\omega_0) \delta(k_x-\hat{k}_x, k_y-\hat{k}_y, k_z-\hat{k}_z),
\end{equation}
which gives a fully localized representation in Fourier space and a non-localized (global) representation in real space. Similar results exist for cylindrical and spherical waves~\cite{Williams1999}. 

In the field of room acoustics, experimental evidence~\cite{Verburg2018} suggests that plane wave expansions can sparsely represent stationary acoustic fields at low frequencies (i.e., the modal region). In this frequency region, it is known that the acoustic field in the room is composed of a few modes, which can be represented by a few plane waves. Similar reports have indicated that monopole sources provide sparse representations of compact sound sources~\cite{Richard2017, Fernandez-Grande2017}, which are fully localized in real space and broadband in the wavenumber domain. 

\subsection{Acoustic radiation in Fourier space}
\label{sec:acoustcones}
A solution of the linearized wave equation of the form described in Eq.~\ref{eq:ansatz} explicitly depends on the wavenumber vector $\textbf{k}$ and frequency $\omega$. The relation between these two quantities is called the dispersion relation, and for non-dispersive media (e.g., air) reads $\omega^2 = c_0^2 |\mathbf{k}|^2$. Plane waves propagate to the far field or decay in the near field (also known as evanescent waves). Consider without loss of generality that a plane wave is propagating or decaying along the $z$-direction. Then, propagating waves are smooth periodic functions defined across the whole of space--time and have acoustic wavelengths $\lambda \geq 2\pi /k_0$. And evanescent waves are smooth functions that decay exponentially away with $z$ and have wavelengths $\lambda < 2\pi /k_0$. While a propagating wave has a real wavenumber along the $z$-direction
\begin{equation}
    k_z = \sqrt{k_0^2-k_x^2-k_y^2},
\end{equation}
an evanescent wave has an imaginary wavenumber in that direction
\begin{equation}
    k_z = i k_z^\prime = i \sqrt{k_x^2 + k_y^2 - k_0^2}. 
\end{equation}
Substituting the above into \eqref{eq:ansatz} gives a propagating plane wave
\begin{equation}
    \hat{p}(\mathbf{r},\omega_0) = P(\omega_0) e^{i(k_x x + k_y y + k_z z)}
\end{equation}
and an evanescent plane wave
\begin{equation}
    \hat{p}(\mathbf{r},\omega_0) = P(\omega_0) e^{i(k_x x + k_y y)} e^{-k_z^\prime z},
\end{equation}
respectively.

\begin{figure}[!ht]
    \centering
    \includegraphics[width=0.9\linewidth,trim=0cm 0cm 0cm 0cm, clip]{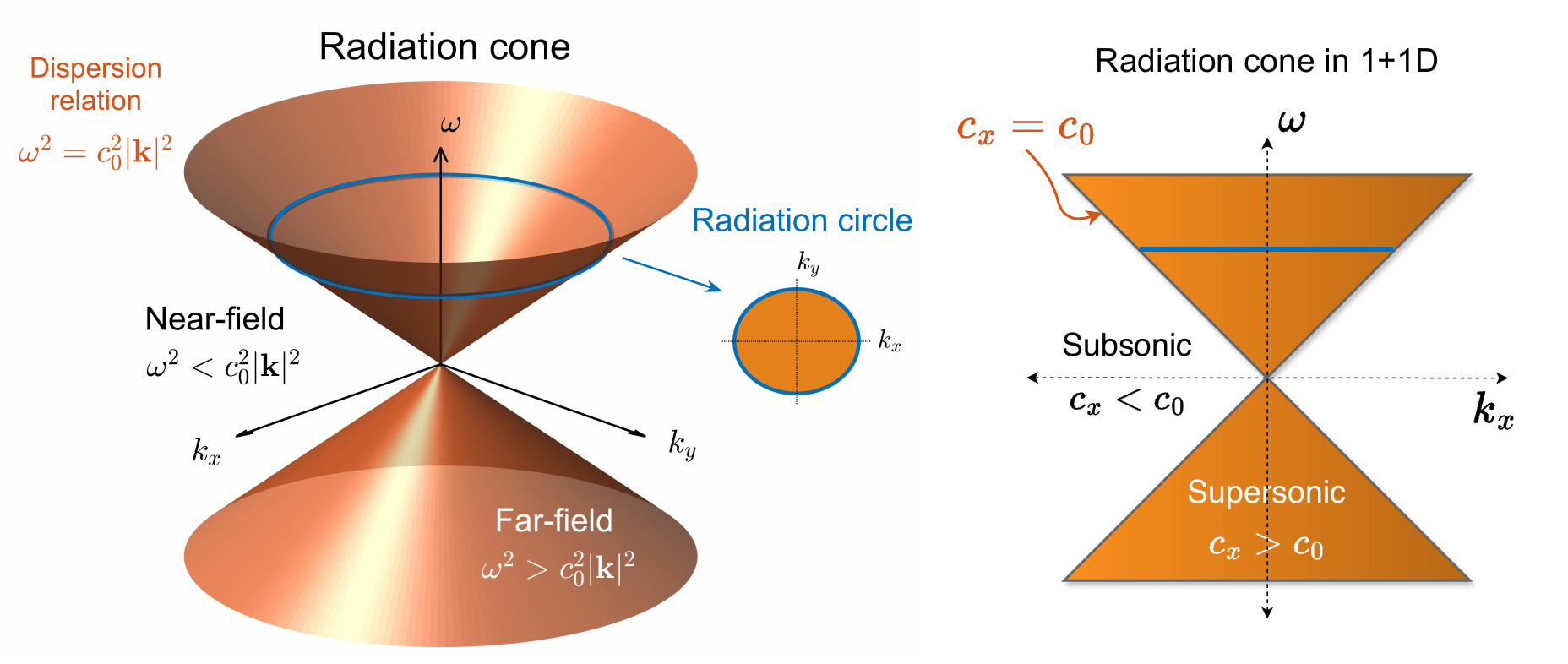}
    \caption{Left: The radiation cone divides acoustic waves into far-field and near-field waves. The radiation circle follows when considering stationary wave solutions, i.e., the Helmholtz equation~\eqref{eq:helmholtz}. Right: Projection of the radiation cone in 1+1D, showing the cone regions with supersonic (far-field) and subsonic (near-field) trace waves, i.e., phase velocities greater and smaller than $c_0$, respectively. The cone boundary corresponds to a phase velocity equal to $c_0$. Cones arise naturally in special relativity due to the principle of causality: a particle inside the cone remains inside the cone. Using the acoustics analogy, a far-field wave inside the radiation cone remains a far-field wave.}
    \label{fig:cone}
\end{figure}

For all values of $(k_x,k_y) \in \mathbb{R}^2$ and a specific frequency $\omega_0$, propagating and evanescent waves are thus divided by the so-called radiation circle~\cite{Williams1999}. As illustrated in Figure~\ref{fig:cone}, the extrusion of the radiation circle as the frequency $\omega \in \mathbb{R}$ varies results in the radiation cone. The presence of this cone suggests that a natural decomposition of the wavefield must preserve this cone. 

\subsection{Trace waves and phase velocities}

While waves of the form in Eq.~\eqref{eq:ansatz} propagate in three dimensions, we will consider one-dimensional measurements of these wavefields (such as those obtained using a linear microphone array) in this work. The result is a projection of the wavefield along a single spatial coordinate, which induces a one-dimensional wavefield along that coordinate.

From Fourier acoustics theory~\cite{Williams1999}, it is known that the projection of such an acoustic wave along one spatial coordinate, say $x$, will result in a so-called \emph{trace wave}
\begin{equation}
    p(x,t) = \hat{p} e^{i (k_x x - \omega_0 t)}
\end{equation}
with amplitude $\hat{p} \in \mathbb{C}$, which propagates with a phase velocity $c_x = \omega_0/k_x$ and has an associated trace wavelength $\lambda_x = 2\pi/k_x$ and trace wavenumber $k_x$ along the $x$-axis. In the remainder of this paper, we shall denote the trace wavenumber along the $x$-axis simply with $k$. 

Let us consider, without loss of generality, a propagating wave with azimuth angle $\alpha = 0$ and some elevation angle $\beta$ with respect to the normal of the $x$-axis. Then, using~\eqref{eq:kxkykz}, the trace wavenumber $k = k_0 \sin \beta$, and the phase velocity along the $x$-axis reads $c_x = c_0/ \sin \beta$. On the one hand, for instance, a propagating wave with $\beta = \pi/2$ is recorded at a line of receivers on the $x$-axis with a phase velocity $c_x = c_0$. Similarly, if $\beta = 0$, the propagating wave has infinite phase velocity. On the other hand, an evanescent wave has an imaginary $\beta$, which results in a phase velocity always smaller than the sound speed $c_0$ and an exponential decay along the normal of the $x$-axis. In other words, as illustrated on the right of Figure~\ref{fig:cone}, propagating waves are inside the radiation cone, with phase velocity $c_x > c_0$, and are commonly referred to as supersonic waves. Conversely, evanescent waves are outside the radiation cone, with phase velocity $c_x < c_0$, and are commonly referred to as subsonic waves. 

\subsection{Encoding the dispersion relation with the Poincar{\'e} group}
\label{sec:dispoincare}
From a geometric perspective, we have seen in the previous sections that the dispersion relation defines the radiation cone, as shown in Figure~\ref{fig:cone}. Representing a wavefront localized in the wavenumber--frequency domain needs to respect such a geometric structure. In other words, encoding wave propagation physics into the representation system means no transformation of the signal-acquisition method can modify the dispersion relation. This is equivalent to requiring the dispersion relation to be covariant with respect to the group of transformations underlying the signal representation. 

As experimentally shown in Figure~\ref{fig:setup}(c), acoustic fields are localized into phase velocity cones and hyperbolic scales (or wavenumber--frequency dilations). These theoretical and experimental arguments suggest that the Poincar{\'e} group is a natural candidate, as it preserves cone symmetries and much of its mathematical properties have been studied in Poincaré-based wavelet systems~\cite{Ali2000}. Specifically, the orbits of the Lorentz subgroup form hyperbolae in the wavenumber--frequency domain, reproducing the structures observed in experimental data. Concretely, these boosts modify the phase velocity of a wavefield, $c_x$, while preserving its scale. Conversely, the group of (isotropic) dilations modifies the scale of the wavefield while preserving the phase velocity. For these reasons, the boostlet construction is based on the semi-direct product of the Poincar{\'e} group with dilations.

\begin{figure}[!ht]
    \centering
    \includegraphics[width=0.85\linewidth]{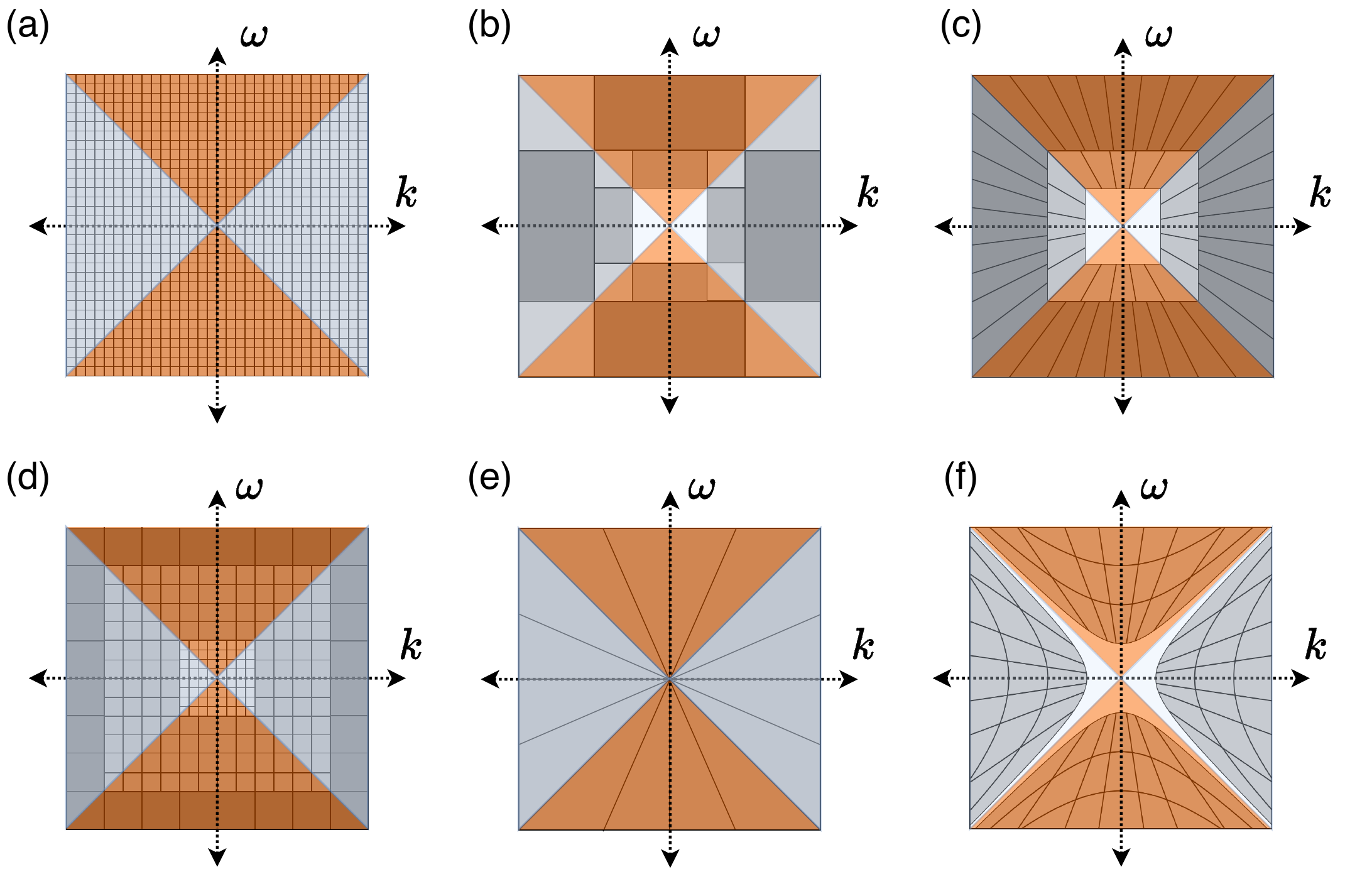}
    \caption{Tessellating the 2D wavenumber--frequency (Fourier) domain with different cell partitions. (a) Plane waves. (b) Isotropic wavelets. (c)  Curvelets~\cite{Candes2004} and shearlets~\cite{Labate2005}. (d) Wave atoms~\cite{Demanet2009}. (e) Directional sub-band filters~\cite{Pinto2010}. (f) Boostlets. The orange cone overlaid corresponds to the radiation cone. The white regions in (b), (c), and (f) denote the wavelet, curvelet/shearlet, and boostlet scaling functions, respectively.}
    \label{fig:tessellate}
\end{figure}

To illustrate the above intuition in a geometric fashion, let us apply various cell partitions to the 2D wavenumber--frequency domain, as shown in Figure~\ref{fig:tessellate}. The radiation cone is overlaid in orange. Plane waves, wavelets, and wave atoms partition the spectrum into rectangular cells of various areas (see Figures~ \ref{fig:tessellate}(a), (b), and (c)). Unless the spectrum has a high enough resolution, most of the coefficients nearest the cone boundary are inside and outside, which does not respect the dispersion relation. While curvelets and shearlets, shown in Figure~\ref{fig:tessellate}(c), can be defined with a conic partition, their decomposition scales (corona) do not follow the hyperbolae localized into phase velocity cones as shown in Figure~\ref{fig:setup}(c). The directional sub-band filters proposed by Pinto and Vetterli~\cite{Pinto2010}, shown in Figure~\ref{fig:tessellate}(e), consist of band-pass filtering transient plane waves within intervals of phase velocities. These filters are closest to boostlets due to the use of phase velocity cones. The boostlet partitioning is shown in Figure~\ref{fig:tessellate}(f), initially proposed in~\cite{Zea2021}, and composed of hyperbolae localized in phase velocity cones and a concave-shaped scaling function (more details in Section~\ref{sec:scaling}). The main difference between boostlets and directional filters lies in the localization of narrow-band frequency content within hyperbolic bands. Localizing narrow-band frequencies in an acoustic wave field allows for a flexible description of frequency-dependent absorption and scattering phenomena, e.g., different bands can have different weights. 

\section{Continuous boostlet transform}

The framework to handcraft the continuous boostlet transform, which follows the same spirit of previous harmonic analysis constructions~\cite{Candes2004,Labate2005,Gorodnitskiy2011}, is presented in this section. The derivations in this paper are in 2D space--time, that is, 1D space and time, and extensions to higher spatial dimensions are the subject of future work. The admissibility condition is derived using the group structure of the transform. Lastly, Parseval's relationship associated with the boostlet transform is shown, leading us to the definition of the boostlet scaling function for finite dilations. 

\begin{definition}[]
Define a dilation matrix $D_a$ and a boost matrix $B_\theta$ acting on space--time vectors $\varsigma = (x,t)^\transp \in \mathbb{R}^2$ as
\begin{equation}
    D_a = \begin{pmatrix} a & 0 \\ 0 & a \end{pmatrix}, 
    \,\,\, B_\theta = \begin{pmatrix} \cosh \theta & -\sinh \theta \\ -\sinh \theta & \cosh \theta \end{pmatrix},
\end{equation}
with dilation parameter $a \in \mathbb{R}^+$, and Lorentz boost (hyperbolic rotation) parameter $\theta \in \mathbb{R}$.
It is often convenient to combine these two transformations into a single dilation--boost matrix $M_{a,\theta}$ given by
\begin{equation}
    M_{a,\theta} = D_a B_\theta = B_\theta D_a = \begin{pmatrix} a \cosh \theta & -a \sinh \theta \\ -a \sinh \theta & a \cosh \theta \end{pmatrix}.
\end{equation}
\end{definition}

\begin{definition}[]
Given a translation vector in 2D space--time $\tau = (\tau_x,\tau_t)^\transp \in \mathbb{R}^2$, a dilation factor $a \in \mathbb{R}^+$, and a boost parameter $\theta \in \mathbb{R}$, we define the boostlet function $\psi_{a,\theta,\tau}(\varsigma) \in L^2(\mathbb{R}^2)$ as
\begin{equation}
    \psi_{a,\theta,\tau}(\varsigma) = a^{-1} \psi \left(D_a^{-1} B_\theta^{-1}(\varsigma-\tau)\right) = a^{-1} \psi\left(M_{a,\theta}^{-1} (\varsigma - \tau)\right),
\end{equation}
respectively, for some mother boostlet $\psi \in L^2(\mathbb{R}^2)$.
\end{definition}

Let us define the Fourier transform of spatiotemporal functions $f(\varsigma)$ as
\begin{equation}
    \hat{f}(\xi) = \int_{\mathbb{R}^2} f(\varsigma) e^{-2\pi i \xi^\transp \varsigma} d\varsigma.
\end{equation}
Here $\xi$ is the wavenumber--frequency vector $\xi = (k, \omega)^\transp$ representing the spectrum of the wavefield in both space and time.
It is then seen that the Fourier transform of the boostlet function is given by
\begin{equation}
    \hat{\psi}_{a,\theta,\tau}(\xi) = a\, e^{-2\pi i \tau^\transp \xi} \hat{\psi}(M_{a,\theta}^\transp \xi).
    \label{eq:FourierBoostlet}
\end{equation}

In the following, we will require $\hat{\psi}(\xi)$ to be supported in the near-field cone, i.e. $\hat{\psi}(\xi) = 0$ for $\xi = (k, \omega)$ such that $|k| < |\omega|$.
This near-field boostlet can then be transformed into a far-field boostlet by defining
\begin{equation}
    \psi^*(x, t) = \psi(t, x).
\end{equation}
This function will then have the property that $\hat{\psi}^*(\xi) = 0$ for all $|k| > |\omega|$, and is thus supported in the far-field.
Since the Poincaré group preserves the radiation cone, we will have that $\psi_{a,\theta,\tau}$ and $\psi_{a,\theta,\tau}^*$ are also supported in the near and far fields, respectively.
Decomposing a wavefield $f(x, t)$ using these boostlets gives us the desired boostlet transform.

\begin{definition}[]
Set 
\begin{equation}
    \mathbb{B} = \{ (a,\theta,\tau) : a \in \mathbb{R}^+, \theta \in \mathbb{R}, \tau \in \mathbb{R}^2 \}.
\end{equation}
For a near-field mother boostlet $\psi \in L^2(\mathbb{R}^2)$, we define the \emph{continuous boostlet transform} of a spatiotemporal function $f(\varsigma) \in L^2(\mathbb{R}^2)$ by the mapping
\begin{equation}
    f \to \mathcal{B}_f = \left( \langle f, \psi_{a, \theta,\tau} \rangle, \langle f, \psi_{a, \theta, \tau}^* \rangle \right)_{(a,\theta,\tau) \in \mathbb{B}}.
\end{equation}
\end{definition}

\begin{remark}
    The boostlet transform expands $f$ into a redundant representation, comprised of its inner products with the boostlet functions $\psi_{a,\theta,\tau}$ and $\psi_{a,\theta,\tau}^*$. From a signal-decomposition perspective, boostlets band-pass filter the signal $f$ at various scales $a\in \mathbb{R}^+$ (wavelength intervals) and boosts $\theta \in \mathbb{R}$ (phase velocity intervals). 
\end{remark}

How the set of boostlet functions respects the dispersion relation (and the acoustic radiation cone) is the subject of the following sections.
In particular, while the above transform can be computed for any choice of functions $\psi \in L^2(\mathbb{R}^2)$, it is not clear whether the original wavefield can be recovered from the transform.
To investigate this, we begin with the definition of the boostlet group, which allows us to derive the admissibility condition for $\psi$ as well as several useful properties of the transform.

\subsection{Boostlet group}

To better understand the properties of the continuous boostlet transform defined above, we first need a more detailed characterization of the group of transformations used to generate the boostlet system. The boostlet group is also known in the wavelet literature as the affine Poincaré group, the Weyl--Poincaré group, and the similitude group~\cite{Ali2000}. The proof of the following lemma is included in~\ref{prf:group}.

\begin{lemma}
\label{lmm:group}
The set $\mathbb{B} = \mathbb{R}^+ \times \mathbb{R} \times \mathbb{R}^2$ equipped with a product $\cdot$ given by
\begin{equation}
    \left( a, \theta, \tau \right) \cdot \left( a', \theta', \tau' \right) = \left( a a', \theta + \theta', \tau + B_\theta D_a \tau' \right)
\end{equation}
forms a group with identity $(1,0,0)$. 
\end{lemma}

We will refer to the group $\mathbb{B}$ with the above structure as the \emph{boostlet group}.
More generally, this can be thought of as the semi-direct product of the Poincaré group with isotropic dilations in space--time~\cite{Ali2000}.
At this point, we find no reason to dilate \textit{space--time anisotropically} because the propagation medium considered in this study is non-dispersive. This can be compared to, for example, curvelets, shearlets, and related constructions, which have strong reasons to dilate \textit{real space parabolically}. Incorporating such parabolic dilations in space may be sensible when extending the boostlet group to higher spatial dimensions. As it will later become evident, however, the action of the boost matrix $B_\theta$ does generate anisotropic elements in space--time.

Let us now consider the action of the boostlet group $\mathbb{B}$ on the space of wavefields, here represented with functions in $L^2(\mathbb{R}^2)$. 
In addition to its group structure, $\mathbb{B}$, as a subset of $\mathbb{R}^4$, also possesses a differentiable manifold structure.
Consequently, $\mathbb{B}$ is a Lie group, which, among other things, lets us define a Haar measure on the group. We include the proofs of the following lemmata in~\ref{prf:unitary} and~\ref{prf:haar} for completeness.

\begin{lemma}
\label{lmm:unitary}
Define $\sigma: \mathbb{B} \to \mathfrak{U}( L^2(\mathbb{R}^2))$, where $\mathfrak{U}( L^2(\mathbb{R}^2))$ denotes the group of unitary operators on $L^2(\mathbb{R}^2)$, as
\begin{equation}
\label{eq:sigma}
    \sigma(a,\theta,\tau)\psi(\varsigma) = \psi_{a,\theta,\tau}(\varsigma) = a^{-1} \psi(D_a^{-1}B_\theta^{-1}(\varsigma-\tau)).
\end{equation}
Then $\sigma$ is a unitary representation of $\mathbb{B}$ on $L^2(\mathbb{R}^2)$.
\end{lemma}

\begin{lemma}
\label{lmm:haar}
    The left-invariant Haar measure of $\mathbb{B}$ is given by
\begin{equation}
    \mathrm{d}\mu(a, \theta, \tau) = a^{-3} \mathrm{d}a \mathrm{d}\theta \mathrm{d}\tau.
\label{eq:haar}
\end{equation}
\end{lemma}

It is known from classical wavelet theory~\cite{Weiss2001} that the existence of a reproducing formula that recovers the original function from the transform relies on the admissibility of the wavelet function $\psi$. In light of this, we define the admissibility of the mother boostlet $\psi$ in the following manner.

\begin{definition}
    Let $\psi \in L^2(\mathbb{R}^2)$ define a continuous boostlet transform $(a, \theta, \tau) \mapsto \langle f, \psi_{a,\theta,\tau}\rangle$.
    We refer to $\psi$ as \emph{admissible} with respect to the boostlet group $\mathbb{B}$, if
    \begin{equation}
        f = \int_{\mathbb{R}^2} \int_{\mathbb{B}} \left( \langle f, \psi_{a,\theta,\tau} \rangle \, \psi_{a,\theta,\tau} + \langle f, \psi_{a, \theta, \tau}^* \rangle \, \psi_{a, \theta, \tau}^* \right)\, \mathrm{d}\mu(a,\theta,\tau) 
    \label{eq:recformula}
    \end{equation}
    for all $f \in L^2(\mathbb{R}^2)$, where $\mathrm{d}\mu(a,\theta,\tau)$ is the left-invariant Haar measure of $\mathbb{B}$ in Eq.~\eqref{eq:haar}.
\end{definition}

A rigorous characterization of these admissible boostlets is given later in Theorem~\ref{thm:admissible}.

\subsection{Admissible boostlets}
\label{sec:admissible}

The next step is now to characterize the set of admissible functions $\psi_{a,\theta,\tau} \in L^2(\mathbb{R}^2)$. Furthermore, this section shows how the dispersion relation emerges from the admissibility condition for continuous boostlets. It is also shown that the 2D continuous boostlet transform is an isometry for $L^2(\mathbb{R}^2)$.  

Let us begin with a lemma that characterizes the square integrability of the boostlet functions and their corresponding admissibility. The proof of this lemma is found in~\ref{prf:admissible}.
\begin{lemma}
\label{lmm:admissible}
Given $f(\varsigma) \in L^2(\mathbb{R}^2)$, a near-field boostlet function $\psi(\varsigma)\in L^2(\mathbb{R}^2)$, and the wavenumber--frequency vector $\xi = (k,\omega) \in \mathbb{R}^2$, it follows that
\begin{equation}
    \int_\mathbb{B} | \langle f,\psi_{a,\theta,\tau} \rangle |^2  + | \langle f, \psi_{a, \theta, \tau}^* \rangle |^2 \frac{\mathrm{d}a\mathrm{d}\theta \mathrm{d}\tau}{a^3} = \int_{\mathbb{R}^2} | \hat{f}(\xi) |^2 \mathrm{d}\xi \int_0^{\infty} \int_{-k'}^{k'} \frac{| \hat{\psi}(k',\omega')|^2}{k'^2-\omega'^2}  \mathrm{d}\omega' \mathrm{d}k'.
\end{equation}
\end{lemma}

The following theorem introduces the admissibility condition associated with the boostlet transform. It is worth noting that the admissibility condition associated with space--time wavelets has previously been studied by Bohnke~\cite{Bohnke1991} as well as Ali, Antoine and Gazeau~\cite{Ali2000}, and here we include our proof for completeness in~\ref{prf:thm:admissible}.

\begin{theorem}
\label{thm:admissible}
Let $\psi \in L^2(\mathbb{R}^2)$ be such that
\begin{equation}
\label{eq:admissboostlet}
    \Delta = \int_0^{\infty} \int_{-k}^{k} \frac{\left|\hat{\psi}(k,\omega)\right|^2}{k^2 - \omega^2} \mathrm{d}\omega \mathrm{d}k = 1.
\end{equation}
Then, $\psi$ is an admissible boostlet in 2D space--time. 
\end{theorem}

We note that the requirement that $\Delta$ equal one in the above result can be replaced by having $\Delta < \infty$ and dividing $\psi$ by $\sqrt{\Delta}$ to obtain an admissible boostlet function~\cite{Ali2000}. Examples of admissible boostlets $\psi \in L^2(\mathbb{R}^2)$ are tensor wavelets defined in the parameter space $(a,\theta) \in \mathbb{R}^+ \times \mathbb{R}$ (shown in Figure~\ref{fig:boostfun}). To illustrate this, let us introduce a curvilinear change of coordinates for the near-field cone:
\begin{equation}
    \label{eq:diffeo}
    \begin{split}
        a(k, \omega) & = \sqrt{k^2 - \omega^2}, \\
        \theta(k, \omega) & = \operatorname{atanh} \frac{\omega}{k}.
    \end{split}
\end{equation}
Given the Fourier transform $\hat{\Psi}(a, \theta)$ of analytical 2D wavelet, we can then define a boostlet through a change of variables:
\begin{equation}
\label{eq:anwavelet}
    \hat{\psi}(k, \omega) = \hat{\Psi}(a(k, \omega), \theta(k, \omega)).
\end{equation}
Inserting this into the formula for $\Delta$, we obtain that
\begin{equation}
    \Delta = \int_0^{\infty} \int_{\mathbb{R}} \frac{|\hat{\Psi}(a, \theta)|^2}{a} \mathrm{d}\theta \mathrm{d}a.
\end{equation}
In other words, we can construct $\hat{\Psi}(a, \theta)$ as the tensor product of an admissible 1D wavelet $\hat{\Psi}_1(a)$ and a lowpass filter $\hat{\Psi}_2(\theta)$ and the resulting boostlet will be admissible after change of variables.
Indeed, this will be the approach for constructing a discrete boostlet frame in Section~\ref{sec:discrete}.

One can then define a Cartesian grid in the wavenumber--frequency domain $(\omega, k)$, apply the mapping~\eqref{eq:diffeo} to said grid and sample the values of, for instance, 2D tensor Meyer wavelets in the $(a,\theta)$ space (where we have a wavelet in the $a$ direction and a scaling function in the $\theta$ direction).

A physical interpretation of Eq.~\eqref{eq:admissboostlet} can be noted. The function emerging in the denominator, $|\omega^2 - k^2|$, corresponds to the Minkowski distance from the acoustic radiation cone in Fourier space, described by the dispersion relation as shown in Figure~\ref{fig:cone}. On the one hand, admissible boostlets inside the cone, i.e., $\omega^2 > k^2$, are far-field waves, and admissible boostlets outside the cone, i.e., $\omega^2 < k^2$, are near-field waves. On the other hand, boostlets are not admissible on the radiation cone, i.e., $\omega^2 = k^2$. 

\subsection{Boostlet scaling function}
\label{sec:scaling}
This section derives the reconstruction formula and the scaling function associated with the boostlet transform when the sets of group parameters are bounded. In many practical applications, even in a continuous fashion, such a constraint is required~\cite{Kutyniok2008}. 

Let us consider the case in which the dilation parameter $a$ is bounded from above by a value $\Omega\in\mathbb R^+$. As a consequence, the validity of Eq.~\eqref{eq:recformula} cannot be guaranteed for a generic function $f\in L^2(\mathbb R^2)$ as such a requirement prevents sampling the full $\mathbb R^2$ space.

\begin{definition}
Let the boostlet group parameters be bounded as $a\in(0,\Omega)$ and $\theta\in\mathbb R$. Then, we can define:
\begin{equation}
\label{eq:delta_omega}
    \Delta^\Omega(\xi) = \int_0^\Omega \int_{\mathbb R}\left|\hat{\psi}(M^\transp_{a,\theta}\xi)\right|^2 + \left|\hat{\psi}^*(M^\transp_{a,\theta}\xi)\right|^2\frac{\mathrm{d}a\mathrm{d}\theta}{a}
\end{equation}
Let $\phi(\varsigma) \in L^2(\mathbb{R}^2)$ such that
\begin{equation}
    |\hat{\phi}(\xi)|^2 + \Delta^\Omega(\xi) = 1
\end{equation}
for all $\xi \in \mathbb{R}^2$.
We call $\phi(\varsigma)$ the \emph{scaling function} associated with the boostlet $\psi(\varsigma)$ at scale $\Omega$.
\end{definition}

Denoting the translation of $\phi(\varsigma)$
\begin{equation}
    \phi_\tau(\varsigma) = \phi(\varsigma - \tau),
\end{equation}
we can now form the inner products $\langle f, \phi_\tau \rangle$ for all $\tau \in \mathbb{R}^2$ in order to capture the parts of the signal ignored by the continuous boostlet transform when restricting it to scales below $\Omega$.
In particular, we have the following theorem.

\begin{theorem}
\label{thm:scaling_function}
Let $f \in L^2(\mathbb{R}^2)$, $\psi \in L^2(\mathbb{R}^2)$ be an admissible boostlet, and $\phi \in L^2(\mathbb{R}^2)$ be its associated scaling function at scale $\Omega$.
Then, we have the following reproducing formula
\begin{equation}
    f(\varsigma) = \int_{\mathbb{R}^2} \langle f, \phi_\tau \rangle \phi_\tau(\varsigma) \mathrm{d}\tau +
    \int_{\mathbb R^2}\int_{\mathbb R}\int_0^\Omega \langle f,\psi_{a,\theta,\tau}\rangle \psi_{a,\theta,\tau}(\varsigma) + \langle f, \psi^*_{a,\theta,\tau} \rangle \psi^*_{a,\theta,\tau}(\varsigma) \frac{\mathrm{d}a\mathrm{d}\theta \mathrm{d}\tau}{a^3}.
\end{equation}

\end{theorem}

The proof is similar to that of Theorem~\ref{thm:admissible} in~\ref{prf:thm:admissible} and we omit it here for brevity.

\begin{remark}
    The continuous boostlet transform defined in Theorem~\ref{thm:scaling_function} is an isometry for $L^2(\mathbb{R}^2)$.
\end{remark}

\section{Discrete boostlet transform}
\label{sec:discrete}

We now introduce a discrete formulation of the boostlet transform based on frame theory. This is achieved by carefully designing a set of boostlet functions in order to satisfy a discrete version of the boostlet admissibility condition. We conclude with a sparsity analysis of discrete boostlets with experimentally measured fields, compared with wavelets, wave atoms, curvelets, and shearlets.

\subsection{Frame theory preliminaries}
\label{sec:frames}

Let us begin with a few basic definitions and properties. In the following, let $f$ be given spatiotemporal data belonging to $L^2(\mathbb{R}^2)$, and $\left\{ \psi_\gamma \right\}_{\gamma\in\Gamma}$ be a set of functions in $L^2(\mathbb{R}^2)$ indexed by the discrete set $\Gamma$. 
We now define operations that take us from a continuous function to a discrete set of coefficients and back using this set of functions.

\begin{definition}
    The \emph{decomposition} or \emph{analysis} of the function $f$ into the sequence of functions $\left\{ \psi_{\gamma} \right\}_{\gamma\in\Gamma}$ is then defined as
    \begin{equation}
        f \mapsto \left\{ c_\gamma \right\}_{\gamma\in\Gamma}, \quad \text{where } c_\gamma =  \langle f, \psi_{\gamma} \rangle.
    \end{equation}
    Here, the coefficients $\left\{ c_\gamma \right\}_{\gamma\in\Gamma}$ are so-called \emph{expansion coefficients} resulting from the inner product between the data $f$ and the system $\left\{ \psi_{\gamma} \right\}_{\gamma\in\Gamma}$.
\end{definition}

\begin{definition}
    Similarly, the \emph{synthesis} of the coefficients $\left\{ c_\gamma \right\}_{\gamma\in\Gamma}$ is defined as
    \begin{equation}
    \label{eq:frameExp}
        \sum_{\gamma\in\Gamma} c_\gamma \psi_{\gamma}(\varsigma).
    \end{equation}
    This is the adjoint of the decomposition defined above.
\end{definition}

We are now ready to introduce frame systems and their properties in the following definitions. 
\begin{definition}
    A collection of functions $\left\{ \psi_{\gamma} \right\}_{\gamma\in\Gamma}$ in $L^2(\mathbb{R}^2)$ is a \emph{frame} if there exist constants $0 < A \leq B < \infty$ such that
    \begin{equation}
        \label{eq:frame-bounds}
        A\| f \|_2^2 \leq \sum_{\gamma\in\Gamma} |c_\gamma|^2 \leq B \| f \|_2^2, \,\, \textrm{for all} \,\, f \in L^2(\mathbb{R}^2).
    \end{equation}
\end{definition}
The above expression states that the energy of the expansion coefficients is lower- and upper-bounded by the \emph{frame bounds} $A$ and $B$.
These bounds determine important special cases of frames, described in the following definition.

\begin{definition}
    A frame $\left\{ \psi_{\gamma} \right\}_{\gamma\in\Gamma}$ is called a \emph{tight frame} if it has bounds $A = B$. 
    It is known as a \emph{Parseval frame} if $A = B = 1$.
\end{definition}

It is known that the synthesis operator of a Parseval frame is the inverse of the analysis operator~\cite{Casazza2013}. In the context of this work, we shall use this property to design the discrete boostlet system.

\subsection{Discrete boostlet frames}

Using the frame formalism in Section~\ref{sec:frames}, we want to apply it in the case where the $\psi_\gamma$s are boostlet functions. Concretely, we will construct a mother boostlet $\psi$ such that $\psi_{a,\theta,\tau}(\varsigma)$ (suitably renormalized) form a frame for certain set of $(a, \theta, \tau)$. Let us start by defining
\begin{equation}
    \psi_{\ell,n,m}(\varsigma) = 2^{-\ell} e^{-|n|\delta} \psi(D_{2^{\ell}}^{-1} B_{n\delta}^{-1} (\varsigma - 2^\ell e^{-|n|\delta} m)),
\end{equation}
where $\ell, n \in \mathbb{Z}$ and $m \in \mathbb{Z}^2$, for some $\delta > 0$.
This is essentially $\psi_{a,\theta,\tau}(\varsigma)$ with $a = 2^{\ell}$, $\theta = n\delta$, and $\tau = 2^\ell e^{-|n|\delta} m$, scaled by a factor $e^{-|n|\delta}$ (this is needed to compensate for the higher sampling rate at higher boost levels).
Since we need both near- and far-field boostlets to capture an entire wavefield, we will define $\{\psi_{\ell,n,m}\}_{\ell,n \in \mathbb{Z},m \in \mathbb{Z}^2}$ to be a \emph{boostlet frame} if the frame bound
\begin{equation}
    A \|f\|^2 \le \sum_{\ell,n,m} |\langle f, \psi_{\ell,n,m} \rangle|^2 + |\langle f, \psi^\ast_{\ell,n,m} \rangle|^2 \le B \|f\|^2
\end{equation}
is satisfied. (Here and in the following, we will suppress the index sets of sums where necessary to avoid cumbersome notation.)

We can now show that such a system forms a frame under a certain condition on the mother boostlet $\psi$.
The proof is found in \ref{prf:thm:frame}.
\begin{theorem}
\label{thm:frame}
    The system $\{\psi_{\ell,n,m}\}_{\ell,n \in \mathbb{Z},m \in \mathbb{Z}^2}$ forms a frame if the support of $\widehat{\psi}(\xi)$ is contained in a disk of diameter one and there are $A, B > 0$ such that
    \begin{equation}
        \label{eq:frame-condition}
        A \le \sum_{\ell,n} |\widehat{\psi}(D_{2^{\ell}} B_{n\delta} \xi)|^2 + |\widehat{\psi}^\ast(D_{2^\ell} B_{n\delta} \xi)|^2 \le B
    \end{equation}
    almost everywhere.
    The frame bounds are then given by $A$ and $B$.
\end{theorem}

We thus wish to find $\psi$ such that
\begin{equation}
    \sum_{\ell,n\in\mathbb{Z}} |\widehat{\psi}(D_{2^{\ell}} B_{n\delta} \xi)|^2 + |\widehat{\psi}^\ast(D_{2^{\ell}} B_{n\delta} \xi)|^2 = 1
\end{equation}
for all $\xi \in \mathbb{R}^2$ outside of a set of measure zero.
Note that we include both $\psi(\varsigma)$ and its flipped version $\psi^\ast(\varsigma)$ in the above sum in order to cover for both the near and far fields.

For simplicity, we will focus on constructing a discrete frame for the near field, noting that a complete frame is then obtained by flipping the axes to create a far-field discrete frame.
We thus seek a near-field mother boostlet $\psi(\varsigma)$ (with Fourier support contained in a disk of diameter one) such that
\begin{equation}
    \label{eq:square-sum-near-field}
    \sum_{\ell,n} |\widehat{\psi}(D_{2^{\ell}} B_{n\delta} \xi)|^2 = 1
\end{equation}
for all $\xi = (k, \omega) \in \mathbb{R}^2$ such that $|k| > |\omega|$.

Following the discussion at the end of Section~\ref{sec:admissible}, we use the mapping in Eq.~\eqref{eq:diffeo} to instead consider our boostlets in scale--boost space, setting
\begin{equation}
    \widehat{\psi}(k, \omega) = \hat{\Psi}(a(k, \omega), \theta(k, \omega)),
\end{equation}
where $\hat{\Psi}$ is some function defined on the half-plane $\mathbb{R}^+\times \mathbb{R}$.
One useful property of this mapping is that we can express a dilated and boosted version of the mother boostlet quite easily.
Indeed, since
\begin{equation}
    \begin{split}
        a( D_{a'} B_{\theta'} (k, \omega)^\transp) &= a' a(k, \omega) \\
        \theta( D_{a'} B_{\theta'} (k, \omega)^\transp) &= \theta(k, \omega) - \theta'
    \end{split}
\end{equation}
we obtain
\begin{equation}
    \widehat{\psi}(D_{2^\ell}B_{n\delta} (k, \omega)^\transp) = \hat{\Psi}(2^\ell a(k, \omega), \theta(k, \omega) - n\delta).
\end{equation}
Our goal is thus to tile the half-plane by dilating and shifting $\hat{\Psi}(a, \theta)$.
This is most easily done by defining $\hat{\Psi}(a, \theta)$ as a tensor product
\begin{equation}
\label{eq:tensorboostlet}
    \hat{\Psi}(a, \theta) = \varphi_1(a) \varphi_2(\theta),
\end{equation}
and ensuring that dilated copies of $\varphi_1(a)$ and shifted copies of $\varphi_2(\theta)$ tile the positive real numbers and the whole real line, respectively.
Indeed, suppose that we have that
\begin{equation}
    \begin{split}
        \sum_{\ell \in \mathbb{Z}} |\varphi_1(2^{\ell} a)|^2 &= 1 \qquad \text{for all } a > 0, \text{and} \\
        \sum_{n \in \mathbb{Z}} |\varphi_2(\theta - n\delta)|^2 &= 1 \qquad \text{for all } \theta \in \mathbb{R}.
    \end{split}
\end{equation}
Then we have that
\begin{equation}
    \sum_{\ell, n} |\hat{\Psi}(2^{\ell} a, \theta - n\delta)|^2 = 1
\end{equation}
for all $(a, \theta) \in \mathbb{R}^+ \times \mathbb{R}$.
Since the mapping $(a(k,\omega), \theta(k,\omega))$ maps the near-field cone in the wavenumber--frequency space onto the half-plane in dilation--boost space, we can thus guarantee that Eq.~\eqref{eq:square-sum-near-field} holds for $|k| > |\omega|$.
To construct the frame, we therefore only need to find the functions $\varphi_1$ and $\varphi_2$ that satisfy the above constraints.

\subsubsection{Dilation wavelets $\varphi_1(a)$}
\label{sect:wavelet_dilation}
The wavelets in this section are based on those in \cite{Hauser2014}, which in turn leverages the framework introduced by Meyer in \cite{meyer_v}. In mathematical terms, we define the dilation wavelet $\varphi_1(a)$ as
\begin{equation}
    \varphi_1(a) =  
    \begin{cases}
    \sin\left( \frac{\pi}{2} \nu\left( 3a - 1 \right) \right), & 1/3 \leq |a| \leq 2/3 \\
    \cos\left( \frac{\pi}{2} \nu\left( \frac{3a}{2} - 1 \right) \right), & 2/3 \leq |a| \leq 4/3 \\
    0, & \text{otherwise},
    \end{cases}
\end{equation}
where we employ the usual auxiliary function
\begin{equation}
    \nu(u) =  
    \begin{cases} 
       0, & \text{if } u < 0, \\ 
       35u^4 - 84u^5 + 70u^6 - 20u^7, & \text{if } 0 \leq u \leq 1, \\
       1, & \text{if } u > 1.
    \end{cases}
\end{equation}

This wavelet function satisfies one important property, namely that
\begin{equation}
    \varphi_1^2(a) + \varphi_1^2(2a) = 1
\end{equation}
for $a \in [1/3, 2/3]$.
As a result, we then have
\begin{equation}
    \sum_{\ell \in \mathbb{Z}} \left|\varphi_{1}(2^{\ell}a)\right|^2 = 1
\end{equation}
for all $a > 0$.

\subsubsection{Boost bumps $\varphi_2(\theta)$}
\label{sect:wavelet_translation}

Another useful property of the functions introduced previously is the symmetry of the auxiliary function $v(u)$.
Indeed, it can be verified that $\nu(1-u) = 1 - \nu(u)$, that is, $\nu(u) + \nu(1-u) = 1$.
We therefore define
\begin{equation}
    \varphi_2(\theta) = \sqrt{\nu(\delta(1-|\theta|))},
\end{equation}
for some $\delta > 0$.

Due to the symmetry property described above, we have that
\begin{equation}
    |\varphi_2(\theta)|^2 + |\varphi_2(\theta - \delta)|^2 = 1
\end{equation}
over the interval $[0, \delta]$.
We therefore obtain that
\begin{equation}
    \sum_{n \in \mathbb{Z}} \left|\varphi_{2}(\theta - n\delta)\right|^2 = 1, \quad \forall \theta \in \mathbb{R},
\end{equation}
as desired.

\subsubsection{Scaling function $\phi$}
The above construction requires us to decompose our signal at all scales $2^{\ell}$, ranging from very small to very large.
As discussed in Section~\ref{sec:scaling}, it is essential to limit the maximum scale depending on the task at hand.
This corresponds to limiting the dilation index $\ell$ to some maximum value $L$, yielding the index set
$\Gamma_1^{(L)} = \{\ell \in \mathbb{Z} : \ell < L\}$.

To recover the remaining energy, we need to decompose the signal using a \emph{scaling function} that captures the remaining scales.
For this, we first define
\begin{equation}
    \phi(a) = \begin{cases} 1, & 0 < |a| \le 1/3, \\ 
    \cos\left(\frac{\pi}{2}v\left(3a - 1\right)\right), & 1/3 < |a| \le 2/3, \\ 
    0, & |a| > 2/3, \end{cases}
\end{equation}
and note that
\begin{equation}
    |\phi(2^{L}a)|^2 + \sum_{l = -\infty}^{L-1} |\varphi_{1}(2^{\ell} a)|^2 = 1
\end{equation}
for all $a > 0$.

We then obtain the scaling function
\begin{equation}
\label{eq:discrete:scaling}
    \phi^{(L)}(k, \omega) = \phi(2^{L} a(k, \omega)).
\end{equation}
Together with the near-field boostlets defined previously, we then have
\begin{equation}
\label{eq:discrete:squares:nf}
    |\phi^{(L)}(\xi)|^2 + \sum_{\ell < L, n \in \mathbb{Z}} |\widehat{\Psi}(D_{2^\ell} B_{n\theta}\xi)|^2 = 1
\end{equation}
for all $\xi = (k, \omega)^\transp$ in the near-field cone.
If we also consider the far-field boostlets $\widehat{\psi}^*_\gamma(\xi)$, we obtain
\begin{equation}
    |\phi^{(L)}(\xi)|^2 + \sum_{\ell < L, n \in \mathbb{Z}} |\widehat{\psi}(D_{2^\ell} B_{n\theta} \xi)|^2 + |\widehat{\psi}^*(D_{2^\ell} B_{n\theta} \xi)|^2 = 1
\end{equation}
for all $\xi \in \mathbb{R}^2$.

\begin{figure}[ht!]
    \centering
    \includegraphics[trim={0cm 0 0cm 0},clip,width=\linewidth]{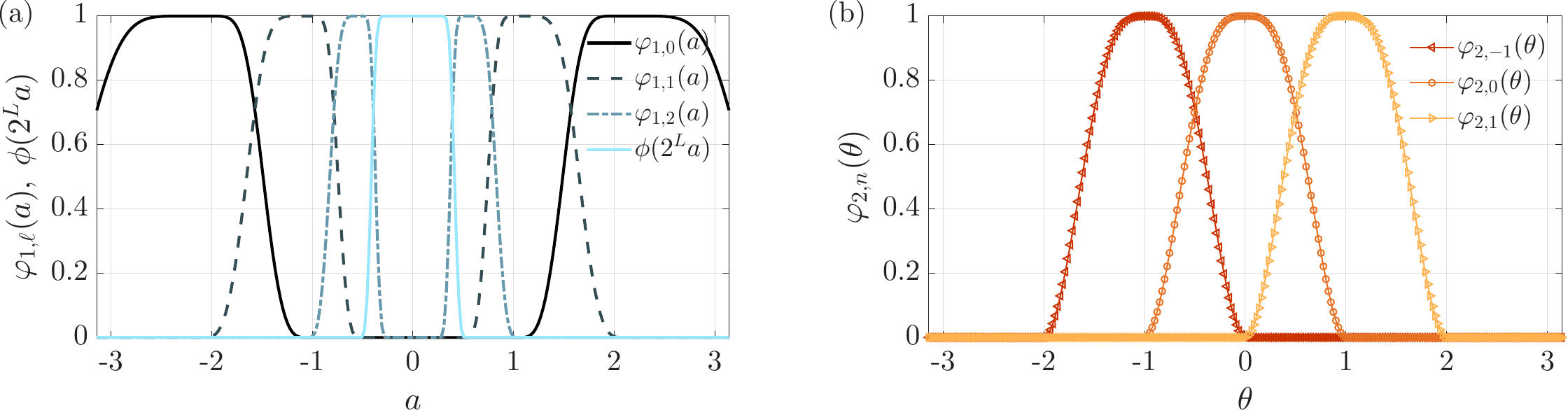}
    \caption{(a) A dilation wavelet family of $L = 3$ scales, $\varphi_{1,\ell}(a)= \varphi_1(2^\ell a)$, with $\ell = \{0, 1, 2\}$, and its corresponding scaling function $\phi(2^L a)$. (b) A boost bump family $\varphi_{2,n}(\theta) = \varphi_2(\theta - n\delta)$, with $n = \{-1, 0, 1\}$ and $\delta = 1$.}
    \label{fig:families}
\end{figure}

Examples of the wavelet family $\varphi_{1,\ell}(a)$ for $\ell = \{0,1,2\}$ and its corresponding scaling function $\phi(2^La)$ for $L = 3$ are shown in Figure~\ref{fig:families}(a). This wavelet family band-pass filters the frequency content of the wavefront into dyadic scales $2^\ell$. The Meyer scaling function filters the low-frequency content of the wavefront at scales larger than $L = 3$, as shown in Eq.~\eqref{eq:discrete:squares:nf}.

The bump family $\varphi_{2,n}(\theta)$ for $n = \{-1,0,1\}$ and $\delta = 1$ is shown in Figure~\ref{fig:families}(b). Each bump band-pass filters the phase speed content of the wavefront with the choice of $\delta$. The family spans a maximum boost, $\theta_{\text{max}}$, linked to the Nyquist sampling rate, say $\omega_{\text{Nyq}}$ in the far-field cone, via $\theta_{\text{max}} = \cosh^{-1}(2^{\ell} \omega_{\text{Nyq}})$. For example, $\theta_{\text{max}} = 2$~rad captures waves traveling with a phase speed $c_x \leq c_0 \tanh{\theta_{\text{max}}} \approx 0.96 c_0$. Waves traveling at greater speeds, $0.96 c_0 < c_x \leq c_0$ , are captured by the boostlet scaling function $\phi^{(3)}(k,\omega)$, which, from Eq.~\eqref{eq:discrete:scaling} is constant along the boost variable. 

\subsection{Visualization of admissible boostlets and wavefield decompositions}
\label{sec:visual}

\begin{figure}[!ht]
    \centering
        \includegraphics[width=\linewidth]{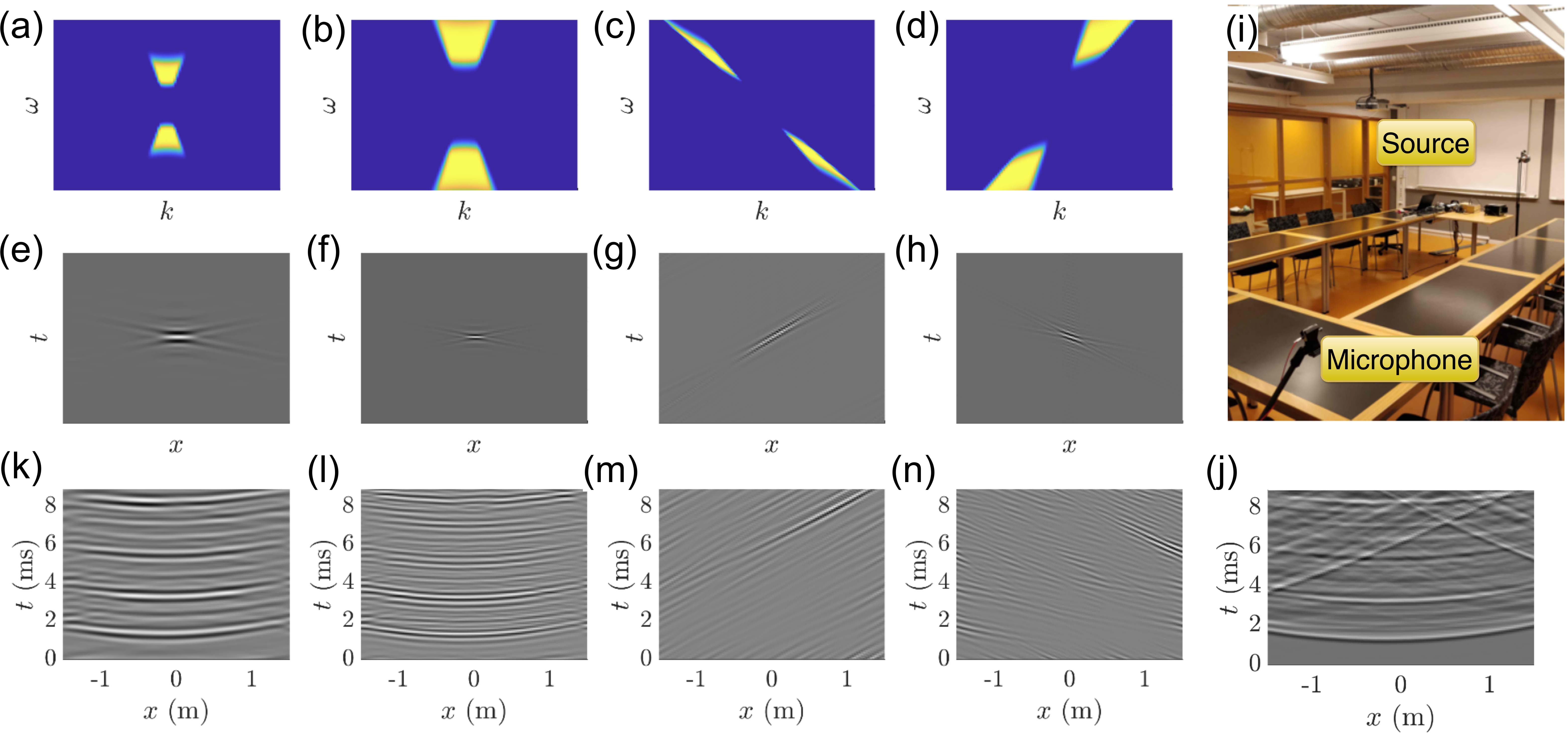}
    \caption{Prototype boostlet functions in (a)-(d) Fourier space and (e)-(h) space--time. (i) Experimental setup of room-acoustical measurement and (j) pressure field measured in space--time. (k)-(n) Convolution between the acoustic field in (j) and the boostlets in (e)-(h).}
        \label{fig:boostfun}
\end{figure}

This section shows examples of admissible boostlets and associated decompositions of a natural acoustic field measured in 2D space--time. Examples of such functions for different wavelengths and phase velocities are shown in Figure~\ref{fig:boostfun}. The convolution between an acoustic field in 2D space--time (see Figures~\ref{fig:boostfun}(i) and (j)) and these boostlets is shown in Figures~\ref{fig:boostfun}(k)-(n). It can be seen that, similar to a shearlet decomposition of the acoustic field~\cite{Zea2019}, boostlets extract (discard) wavefronts aligned (misaligned) with their phase velocity. 

\begin{figure}[!ht]
    \centering
        \includegraphics[width=0.7\linewidth]{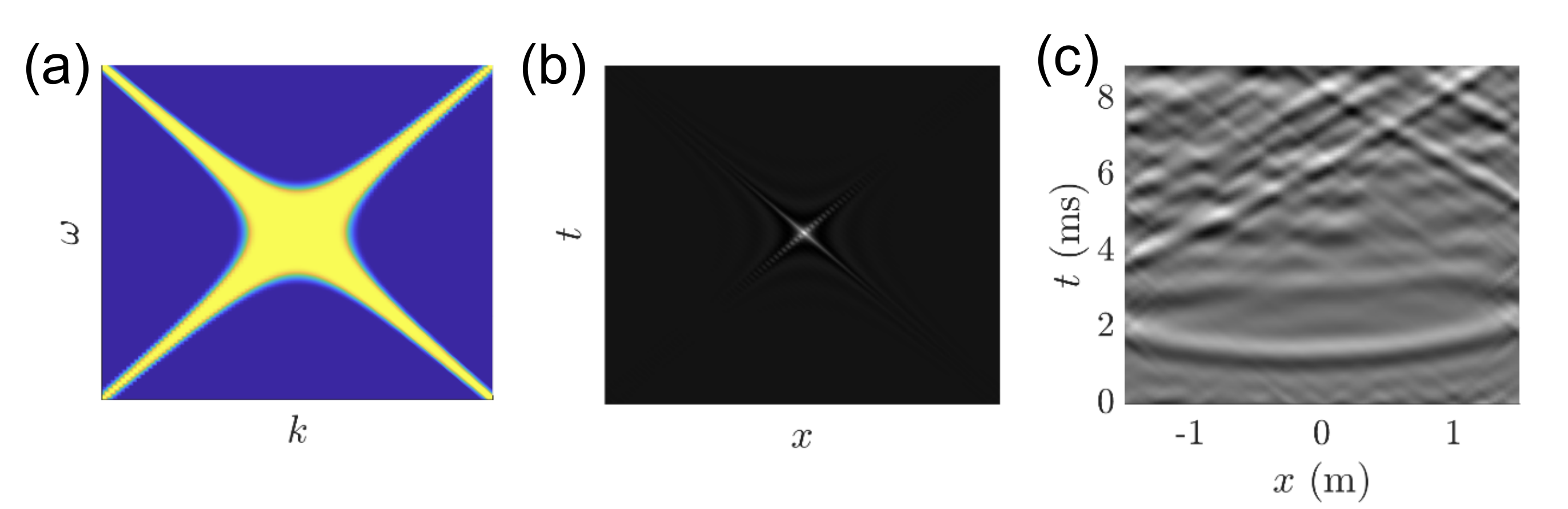}
    \caption{Prototype boostlet scaling function in (a) Fourier space and (b) space--time. (c) Convolution between the acoustic field in Figure~\ref{fig:boostfun}(j) and the scaling function.}
    \label{fig:scalingfun}
\end{figure}

The boostlet scaling function is defined in this study with a Meyer scaling function of $a$ in the $(a,\theta)$ space. The scaling function in Figure~\ref{fig:scalingfun}(a) resembles the wavenumber--frequency spectrum of a transient monopole observed in the theoretical analysis in~\cite{Zea2021}. This can be qualitatively seen with the point-like source localized in space--time in Figure~\ref{fig:scalingfun}(b). Figure~\ref{fig:scalingfun}(c) shows the convolution between the boostlet scaling function and the acoustic field in Figure~\ref{fig:boostfun}(j). It can be noted that the scaling function smoothens the acoustic field. Moreover, the latter exhibits large amplitudes for waves propagating with phase velocities near the speed of sound, in correspondence with the Fourier support in Figure~\ref{fig:scalingfun}(a). 

Let us compare the boostlet and wavelet scaling functions (likewise for curvelets and shearlets) in terms of directional attributes. In the context of acoustic plane waves, directionality typically refers to the (Euclidean) angles between the wavefront and the receiver (cf. $(\alpha,\beta)$ in Eq.~\eqref{eq:ansatz}). This comes from the rationale of stationary wave solutions (i.e., Helmholtz equation). For transient, broadband wave solutions, the phase velocity of the wave, $c_x$, governs its ``directionality'' in space--time. If one were to apply a wavelet decomposition to an acoustic field in space--time, the wavelet scaling function would contain plane waves with all phase velocities whose corresponding lines intersect with a square region of Fourier space (see the white region in Figure~\ref{fig:tessellate}(b)). In contrast, boostlet scaling functions contain plane waves propagating with phase velocities in a concave region of Fourier space (cf. Figure~\ref{fig:scalingfun}(a)). Following a similar reasoning, the wavelet scaling function has a low-pass shape, filtering \emph{coarse} wave content from the acoustic field with all phase velocities. The boostlet scaling function has a combination of (i) a band-pass shape near the cone boundary, filtering \emph{broadband} waves propagating with phase velocity equal to the speed of sound (cf. Figure~\ref{fig:scalingfun}(c)), and (ii) a low-pass shape near the lines $\omega = 0$ and $k=0$, filtering \emph{coarse} waves propagating with all phase velocities. A direct consequence is that the boostlet scaling function contains waves with multiple wavelengths and cannot resolve wavelength scales; that task is left to the boostlets.

\subsection{Sparsity analysis with experimental data}
This section presents an empirical analysis of the magnitude decay of the coefficients when applying the prototype boostlets from Section~\ref{sec:visual} and benchmark systems to acoustic fields measured in space--time. The fields have dimensions $100 \times 100$. The benchmark systems considered are Daubechies45 and Meyer wavelets~\cite{Daubechies1992} (using MATLAB's Wavelet Toolbox), curvelets~\cite{Candes2006b} (using the CurveLab toolbox), wave atoms~\cite{Demanet2007} (using the WaveAtom toolbox), and cone-adapted shearlets~\cite{Hauser2014} (using the FFST toolbox). We set $L = 3$ dilation levels for wavelets, curvelets, shearlets, and boostlets. While Daubechies45 and Meyer wavelets are critically sampled (decimated) with no redundancy, curvelets, wave atoms, shearlets, and boostlets\footnote{Using the 2D tensor wavelet product in Eq.~\eqref{eq:tensorboostlet} and the mapping~\eqref{eq:diffeo}, the redundancy of the boostlet transform as a function of the decomposition scales $L$ follows: $(L\,\text{dilations}) \times (K\, \text{boosts}) \times (2\,\text{cones}) + 1\,\text{scaling function}$. Similar to the rotation and shear levels in curvelets and shearlets, the number of boost levels, here $K = 7$, is independent of $L$ and can be chosen as a trade-off between tessellating Fourier space and avoiding too redundant a representation. Here, the number of boost levels is kept the same for all decomposition scales.} have redundancy factors of $2.5$, $32$, $29$, and $43$, respectively. 

\begin{figure}[!ht]
    \centering
    \includegraphics[width=\linewidth,trim=3cm 0.4cm 3cm 0cm, clip]{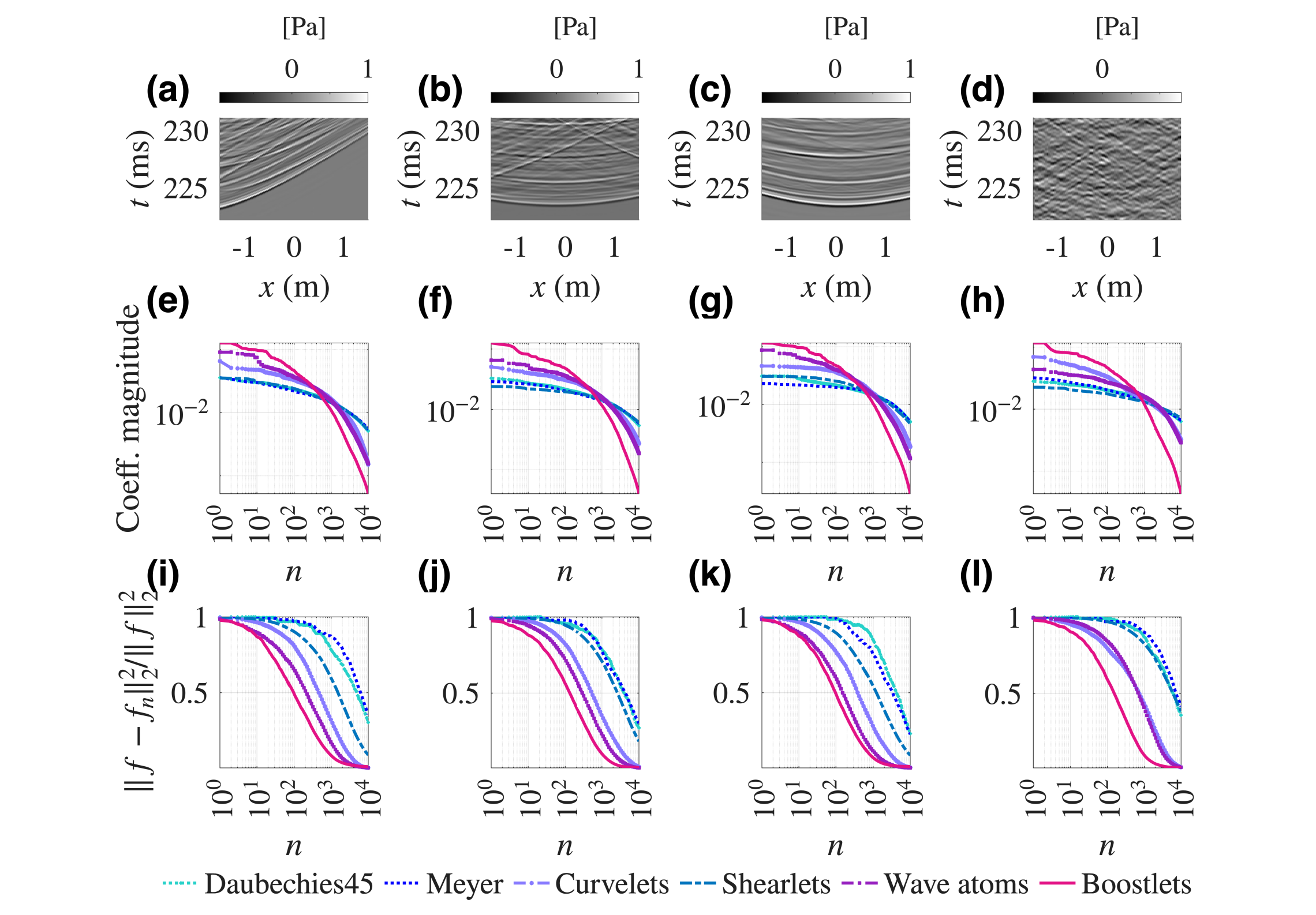}
    \caption{Sparsity analysis applying various expansions to three measured acoustic fields of space--time dimensions $100 \times 100$. (a)-(d): Acoustic fields in 2D space--time from experiments. (e)-(h): Magnitude of the $10\,000$~largest coefficients of Daubechies45 wavelets, Meyer wavelets, curvelets, shearlets, wave atoms, and boostlets sorted in decreasing amplitude. (i)-(l): Relative reconstruction error $\|f - f_n\|_2^2 / \|f\|_2^2$ using up to $10\,000$~largest coefficients.}
    \label{fig:7}
\end{figure}

The analysis is as follows. First, each transform method decomposes the acoustic field to obtain its corresponding coefficients. Next, the coefficients are sorted in descending amplitude, and we examine the sparsity and reconstruction performance using the first $10\,000$~largest coefficients. Similar to the paper by Candès et al.~\cite{Candes2006b}, we compensate for the redundancy of the transforms by normalizing the coefficients by the $\ell_2$-norm of the overall signal to preserve its energy. Besides calculating the coefficient amplitudes to examine their decay, we also compute the corresponding $n$-term relative mean-squared reconstruction error $\|f - f_n\|_2^2 / \|f\|_2^2$, where $f_n$ is the $n$-term approximation of the original wavefield $f$. The results are summarized in Figure~\ref{fig:7}, for four acoustic field examples in (a)-(d), their corresponding $N$-term coefficient decays in (e)-(h), and $N$-term relative reconstruction errors in (i)-(l). 
The code to reproduce the results is in the GitHub repository: \href{https://github.com/eliaszea/Boostlets_SparsityAnalysis}{https://github.com/eliaszea/Boostlets\_SparsityAnalysis}. 

We first note that the acoustic fields in Figures~\ref{fig:7}(a)-(c) are representative of the early part of the microphone recordings (i.e., direct sound and early reflections). In contrast, the acoustic field in Figure~\ref{fig:7}(d) is measured at a later time, having much more significant wave interference and reverberation. For such late reverberation, the sound field becomes increasingly diffuse (i.e., waves propagate in all directions) due to the complex superposition of reflected and scattered waves from walls and objects inside the room. We will see that the sparsity of the representations demonstrates this nuanced effect with an increase of the $\ell_1$-norms.  

For the acoustic fields considered, Figures~\ref{fig:7}(e)-(h) show the magnitudes of the $10\,000$ largest boostlet coefficients decay faster than those of the benchmarks. Wave atoms and curvelet coefficients decay similarly, with curvelets having a sharper decay above $10\,000$ coefficients. Shearlets, Daubechies45, and Meyer wavelets have an asymptotically similar decay. In particular, the boostlet coefficients begin to decay significantly faster than the other expansions above $1\,000$~coefficients. 

Similarly, Figures~\ref{fig:7}(i)-(l) demonstrate that the relative reconstruction errors for these four acoustic fields obtained with boostlets decay consistently faster than the errors obtained with the benchmark systems. There is some variability that appears to depend on the acoustic field itself. For example, the relative errors obtained with wave atoms lie much closer to the errors obtained with boostlets in Figure~\ref{fig:7}(k), compared with Figures~\ref{fig:7}(i), (j), and (l). Furthermore, as shown in Figure~\ref{fig:7}(l) at a later time, the errors obtained with wave atoms and curvelets are very similar, nonetheless with a significantly slower decay than the errors obtained with boostlets. Shearlets, Daubechies45 and Meyer wavelets stand behind with the slowest error decays.  

To complement these observations, the $\ell_1$-norms of the $10\,000$~largest coefficients and the relative MSE errors using the $1\,000$~largest coefficients are obtained with the acoustic fields in Figures~\ref{fig:7}(a)--(d). These metrics are summarized in Table~\ref{tab:1} below. For the acoustic fields considered, the smaller $\ell_1$-norm values of the boostlet coefficients are strong indications of sparsity. This result is further corroborated with the reconstruction errors in the right-most column of the Table, indicating that boostlets attain significantly lower errors than the benchmark systems. 
\begin{table}[!ht]
\centering
\begin{tabular}{lcccc|cccc}
\hline\hline
Method & \multicolumn{4}{c|}{$\ell_1$ (top $10\,000$)} & \multicolumn{4}{c}{Rel.\ error, \% (top $1\,000)$} \\
       & \ref{fig:7}(a)   & \ref{fig:7}(b)   & \ref{fig:7}(c)   & \ref{fig:7}(d)   & \ref{fig:7}(a)     & \ref{fig:7}(b)     & \ref{fig:7}(c)     & \ref{fig:7}(d)     \\ 
\hline
Daubechies45 & 91.5  & 93.1  & 92.5  & 95.2  & 77.4\%  & 79.1\%  & 85.1\%  & 87.3\%  \\ 
Meyer        & 92.9  & 94.5  & 92.6  & 95.4  & 87.2\%  & 77.1\%  & 75.2\%  & 90.4\%  \\ 
Curvelets    & 73.8  & 79.9  & 72.6  & 82.2  & 31.7\%  & 35.0\%  & 27.7\%  & 43.3\%  \\ 
Shearlets    & 91.8  & 94.7  & 89.2  & 97.3  & 59.6\%  & 71.1\%  & 54.3\%  & 82.7\%  \\ 
Wave atoms   & 66.4  & 72.1  & 61.9  & 83.3  & 16.9\%  & 22.8\%  & 11.4\%  & 40.5\%  \\ 
Boostlets    & \textbf{44.9} & \textbf{46.8} & \textbf{41.8} & \textbf{47.0} 
             & \textbf{8.4\%} & \textbf{9.8\%} & \textbf{6.9\%} & \textbf{9.7\%} \\
\hline
\end{tabular}
\caption{Left: $\ell_1$‐norm $\|f_n\|$ of the largest $n = 10\,000$ coefficients for each method and acoustic field (Figures~\ref{fig:7}(a)--(d)). Right: Reconstruction error using the largest $n = 1\,000$ coefficients. Smallest values in each column are in bold.}
\label{tab:1}
\end{table}

It is worth stressing that the prototype boostlets used for the expansions in this section, including those shown in Figure~\ref{fig:boostfun}, are implemented as a frame system with tensor Meyer wavelets using the mapping in Eq.~\eqref{eq:diffeo}. Sparser decompositions may be attainable. Although the results shown are empirical, it is promising that boostlet decompositions have a faster coefficient decay and a smaller reconstruction error from partial coefficients than the benchmark systems. 

\section{Conclusion}

We introduce a boostlet transform for wave-based acoustic signal processing in 2D space--time and nondispersive media.
Several interesting physical interpretations concerning admissibility conditions and the transform parameters are discussed. 
We provide a discrete formulation of boostlet frames using analytical tensor Meyer wavelets in the dilation-boost domain. 
A sparsity analysis with experimentally measured acoustic fields shows that discrete boostlet coefficients exhibit a significantly faster coefficient decay and a significantly lower reconstruction error compared to various benchmarks, including wavelets, curvelets, and wave atoms. 
Ongoing efforts focus on extending the transform to higher spatial dimensions and dispersive media, and formalizing its associated uncertainty principles. 

\section*{Acknowledgments}
E.Z. is financially supported by the Swedish Research Council (Vetenskapsr{\aa}det) under Grant Agreement No. 2020-04668. M.L. is partly supported by the Swedish Research Council (Vetenskapsr{\aa}det) under Grant Agreement No. 2022-03032. The authors would like to thank D. Labate, F. Lizzi, U.P. Svensson, E. Fernandez-Grande, O. Öktem, and C.E. Yarman for many insightful discussions. 

\bibliographystyle{elsarticle-num.bst}


\appendix
\section{Proof of Lemma~\ref{lmm:group}}
\label{prf:group}
It can be checked that $(1,0,0)$ is the identity element. The inverse of an element $(a,\theta,\tau) \in \mathbb{R}^+ \times \mathbb{R} \times \mathbb{R}^2$ is given by
\begin{equation}
    (a,\theta,\tau)^{-1} = \left( \frac{1}{a}, -\theta, -D_a^{-1}B_\theta^{-1} \tau \right),
\end{equation}
since 
\begin{equation}
    (a,\theta,\tau) \cdot \left( \frac{1}{a}, -\theta, -D_a^{-1}B_\theta^{-1} \tau \right) = \left( a\frac{1}{a}, \theta - \theta, \tau - B_\theta D_a D_a^{-1} B_\theta^{-1} \tau \right) = (1,0,0)
\end{equation}
and
\begin{equation}
    \left( \frac{1}{a}, -\theta, -D_a^{-1}B_\theta^{-1} \tau \right) \cdot  (a,\theta,\tau) = \left( \frac{1}{a}a, -\theta + \theta, -D_a^{-1} B_\theta^{-1} \tau + B_{-\theta} D_{\frac{1}{a}} \tau \right) = (1,0,0).
\end{equation}
The product $\cdot$ is also associative, which is shown by
\begin{equation}
    \begin{split}
    ( (a,\theta,\tau) & \cdot(a',\theta',\tau') ) \cdot (a'',\theta'',\tau'') = (a a', \theta+\theta', \tau + B_\theta D_a \tau') \cdot (a'',\theta'',\tau'') \\
    & = (a a' a'', \theta + \theta' + \theta'', \tau + B_\theta D_a \tau' + B_{\theta+\theta'}D_{a a'}\tau'') \\
    & = ( a(a' a''), \theta + (\theta'+\theta''), \tau+B_\theta D_a (\tau' + B_{\theta'}D_{a'} \tau'' ) ) \\
    & = (a,\theta,\tau) \cdot (a'a'',\theta'+\theta'',\tau'+B_{\theta'}D_{a'}\tau'') \\
    & = (a,\theta,\tau) \cdot ( (a',\theta',\tau') \cdot (a'',\theta'',\tau'') ),
    \end{split}
\end{equation}
where we have used the fact that $B_{\theta+\theta'} = B_\theta B_{\theta'}$ which follows from basic trigonometric identities.

\section{Proof of Lemma~\ref{lmm:unitary}}
\label{prf:unitary}
It can be easily checked that the action defined above satisfies the criteria for a group action on $L^2(\mathbb{R}^2)$ and thus forms a representation of $\mathbb{B}$.
Now let us show that this representation is unitary.
Let us start by applying the inverse of $\sigma$, namely $\sigma^{-1}$, as
\begin{equation}
  \sigma^{-1}(a,\theta,\tau) \psi(\varsigma)=a\psi(D_a B_\theta\varsigma+\tau).
\end{equation}
The representation $\sigma$ is unitary if the following relation holds for functions $f,g\in L^2(\mathbb R^2)$:
\begin{equation}
  \langle \sigma f, g\rangle = \langle f, \sigma^*g \rangle = \langle f, \sigma^{-1}g\rangle,
\end{equation}
where $\sigma^{\dagger}$ is the Hermitian conjugate of $\sigma$. The first term is defined as
\begin{equation}\label{eq:ryanair1}
  \langle \sigma f, g \rangle = \int a^{-1} f\left(B_\theta^{-1}D_a^{-1}(\varsigma-\tau)\right)g(\varsigma)\,\mathrm{d}\varsigma,
\end{equation}
while the last term is defined as
\begin{equation}\label{eq:ryanair2}
  \langle f,\sigma^{-1}g \rangle = \int a f(\varsigma) g\left(D_a B_\theta \varsigma+\tau\right)\,\mathrm{d}\varsigma.
\end{equation}
Using the change of variables $\varsigma^\prime=D_a B_\theta \varsigma+ \tau\rightarrow \varsigma=B_\theta^{-1}D_a^{-1}(\varsigma^\prime-\tau)$, and $\mathrm{d}\varsigma = |B_\theta^{-1}D_a^{-1}|\mathrm{d}\varsigma^\prime = a^{-2} \mathrm{d}\varsigma^\prime$ it follows that
\begin{equation}
  \langle \sigma f,g\rangle=\langle f,\sigma^{-1}g\rangle = \int a^{-1} f\left(B_\theta^{-1}D_a^{-1}(\varsigma^\prime-\tau)\right)g(\varsigma^\prime)\,\mathrm{d}\varsigma^\prime\,.
\end{equation}

\section{Proof of Lemma~\ref{lmm:haar}}
\label{prf:haar}
    The left action of the boostlet group $\mathbb{B}$ on the parameters $(a,\theta,\tau)$ follows
    \begin{equation}
        \begin{split}
            a^{\prime\prime} & = a^{\prime}a, \\
            \theta^{\prime\prime} & = \theta^{\prime} + \theta, \\
            \tau^{\prime\prime} & = \tau^{\prime} + B_{\theta^{\prime}} D_{a^{\prime}} \tau.
        \end{split}
    \end{equation}
    Next is to compute the determinant of the Jacobian matrix: 
    \begin{equation}
            \begin{vmatrix}
        \frac{\partial a^{\prime\prime}}{\partial a} & \frac{\partial a^{\prime\prime}}{\partial \theta} & \frac{\partial a^{\prime\prime}}{\partial \tau_x} & \frac{\partial a^{\prime\prime}}{\partial \tau_y} \\
        \frac{\partial \theta^{\prime\prime}}{\partial a} & \frac{\partial \theta^{\prime\prime}}{\partial \theta} & \frac{\partial \theta^{\prime\prime}}{\partial \tau_x} & \frac{\partial \theta^{\prime\prime}}{\partial \tau_y}
        \\
        \frac{\partial \tau_x^{\prime\prime}}{\partial a} & \frac{\partial \tau_x^{\prime\prime}}{\partial \theta} & \frac{\partial \tau_x^{\prime\prime}}{\partial \tau_x} & \frac{\partial \tau_x^{\prime\prime}}{\partial \tau_y}
        \\
        \frac{\partial \tau_y^{\prime\prime}}{\partial a} & \frac{\partial \tau_y^{\prime\prime}}{\partial \theta} & \frac{\partial \tau_y^{\prime\prime}}{\partial \tau_x} & \frac{\partial \tau_y^{\prime\prime}}{\partial \tau_y}
        \end{vmatrix} = 
        \begin{vmatrix}
        a^\prime & 0 & 0 & 0 \\
        0 & 1 & 0 & 0 \\
        0 & 0 & a^\prime \cosh{\theta^{\prime}} & -a^\prime \sinh{\theta^{\prime}}
        \\
        0 & 0 & -a^\prime \sinh{\theta^{\prime}} & a^\prime \cosh{\theta^{\prime}} 
        \end{vmatrix} = {a^\prime}^3.      
    \end{equation}

\section{Proof of Lemma~\ref{lmm:admissible}}
\label{prf:admissible}
Given the Fourier transform of $\psi_{a,\theta,\tau}(\varsigma)$ in Eq.~\eqref{eq:FourierBoostlet} and defining $\widetilde{\psi}_{a,\theta,\tau}(\varsigma)$ as the time-reversal $\psi_{a,\theta,\tau}(-\varsigma)$ of $\psi_{a,\theta,\tau}(\varsigma)$, we obtain, using Plancherel's formula, that
\begin{equation}
\begin{split}
    \int_\mathbb{B} \left| \langle f, \psi_{a,\theta,\tau} \rangle \right|^2 \frac{\mathrm{d}a \mathrm{d}\theta \mathrm{d}\tau}{a^3} & = \int_\mathbb{B} | f \ast \widetilde{\psi}_{a,\theta,0}(\tau) |^2 \frac{\mathrm{d}a\mathrm{d}\theta\mathrm{d}\tau}{a^3}  \\ 
    & = \int_{\mathbb{R}^2} \int_\mathbb{R} \int_0^\infty |\hat{f}(\xi) |^2 \left| \widehat{\widetilde{\psi}_{a,\theta,0}}(\xi) \right|^2 \frac{\mathrm{d}a\mathrm{d}\theta}{a^3} \mathrm{d}\xi \\
    & = \int_{\mathbb{R}^2} \int_\mathbb{R} \int_0^\infty |\hat{f}(\xi) |^2 \left| \hat{\psi}(M_{a,\theta}^\transp \xi)\right|^2 \frac{\mathrm{d}a\mathrm{d}\theta}{a} \mathrm{d}\xi  \\
    & = \int_{\mathbb{R}^2} |\hat{f}(\xi) |^2 \Delta_\psi(\xi) \, \mathrm{d}\xi,
\end{split}
\end{equation}
where 
\begin{equation}
\begin{split}
    \Delta_\psi(\xi) & = \int_\mathbb{R} \int_0^\infty \left| \hat{\psi}(M_{a,\theta}^\transp \xi)\right|^2 \frac{\mathrm{d}a\mathrm{d}\theta}{a} \\ 
    & = \int_\mathbb{R} \int_0^\infty \left| \hat{\psi}(a(k \cosh\theta -\omega\sinh\theta), a(-k\sinh\theta + \omega\cosh\theta))\right|^2 \frac{\mathrm{d}a \mathrm{d}\theta}{a}.
\end{split}
\end{equation}
Introducing the change of variables $k' = a(k\cosh\theta -\omega\sinh\theta)$, and $\omega' = a(-k\sinh\theta + \omega\cosh\theta)$, the Jacobian gives $\mathrm{d}k' \mathrm{d}\omega' = a^{-1}|k'^2-\omega'^2| \mathrm{d}a \mathrm{d}\theta$. To determine the integration area, we note that $\xi$ can be in one of the four cones defined by $\mathcal{C}_+^\mathrm{near} = \{(k, \omega) : k > 0, |\omega| < k\}$, $\mathcal{C}_-^\mathrm{near} = \{(k, \omega): k < 0, |\omega| < -k\}$, $\mathcal{C}_+^\mathrm{far} = \{(k, \omega): \omega > 0, |k| < \omega\}$ and $\mathcal{C}_-^\mathrm{far} = \{(k, \omega): \omega < 0, |k| < -\omega\}$.
Let us denote this cone by $\mathcal{C}(\xi)$.
Since the dilations and boosts preserve these cones, the variables $(k', \omega')$ will span the entirety of the cone $\mathcal{C}(\xi)$ to which $\xi$ belongs.
Thus, it follows that
\begin{equation}
    \Delta_\psi(\xi) = \int_{\mathcal{C}(\xi)} \frac{| \hat{\psi}(k',\omega')|^2}{|k'^2-\omega'^2|} \mathrm{d}k' \mathrm{d}\omega'.
\end{equation}
Since we require $\psi(\varsigma)$ to be in the near field, this integral is zero for $\xi \in \mathcal{C}^\mathrm{far} = \mathcal{C}_-^\mathrm{far} \cup \mathcal{C}_+^\mathrm{far}$, while for $\xi \in \mathcal{C}^\mathrm{near} = \mathcal{C}_-^\mathrm{near} \cup \mathcal{C}_+^\mathrm{near}$, it is constant.
More specifically, we have
\begin{equation}
    \Delta_\psi(\xi) = \begin{cases} 0, & \xi \in \mathcal{C}^\mathrm{far}, \\ \int_{\mathcal{C}_+^\mathrm{near}} \frac{| \hat{\psi}(k',\omega')|^2}{k'^2-\omega'^2} \mathrm{d}k' \mathrm{d}\omega', & \xi \in \mathcal{C}^\mathrm{near}, \end{cases}
\end{equation}
where we have used the fact that $\hat{\psi}(\xi)$ has Hermitian symmetry and therefore satisfies $|\hat{\psi}(-\xi)|^2 = |\hat{\psi}(\xi)|^2$ so we can replace the $\mathcal{C}_-^\mathrm{near}$ with $\mathcal{C}_+^\mathrm{near}$ if necessary.

A similar argument shows that, for the far-field boostlet $\psi^*(\varsigma)$, we have that
\begin{equation}
    \int_\mathbb{B} \left| \langle f, \psi^*_{a,\theta,\tau} \rangle \right|^2 \frac{\mathrm{d}a \mathrm{d}\theta \mathrm{d}\tau}{a^3} = \int_{\mathbb{R}^2} |\hat{f}(\xi) |^2 \Delta_{\psi^*}(\xi) \, \mathrm{d}\xi,
\end{equation}
where
\begin{equation}
    \Delta_{\psi^*}(\xi) = \begin{cases} \int_{\mathcal{C}_+^\mathrm{far}} \frac{| \hat{\psi}^*(k',\omega')|^2}{\omega'^2-k'^2} \mathrm{d}k' \mathrm{d}\omega', & \xi \in \mathcal{C}^\mathrm{far}, \\ 0, & \xi \in \mathcal{C}^\mathrm{near}, \end{cases},
\end{equation}
which can be rewritten as
\begin{equation}
    \Delta_{\psi^*}(\xi) = \begin{cases} \int_{\mathcal{C}_+^\mathrm{near}} \frac{| \hat{\psi}(k',\omega')|^2}{k'^2-\omega'^2} \mathrm{d}k' \mathrm{d}\omega', & \xi \in \mathcal{C}^\mathrm{far}, \\ 0, & \xi \in \mathcal{C}^\mathrm{near}, \end{cases},
\end{equation}
using the definition $\psi^*(x, t) = \psi(t, x)$.

We thus have that
\begin{equation}
    \Delta_\psi(\xi) + \Delta_{\psi^*}(\xi) = \int_{\mathcal{C}_+^\mathrm{near}} \frac{| \hat{\psi}(k',\omega')|^2}{k'^2-\omega'^2} \mathrm{d}k' \mathrm{d}\omega' = \int_0^{\infty} \int_{-k'}^{k'} \frac{|\hat{\psi}(k', \omega')|^2}{k'^2 - \omega'^2} \mathrm{d}\omega' \mathrm{d}k',
\end{equation}
for almost all $\xi \in \mathbb{R}^2$. 

\section{Proof of Theorem~\ref{thm:admissible}}
\label{prf:thm:admissible}
    Plugging in \eqref{eq:admissboostlet} into the identity given by Lemma~\ref{lmm:admissible}, we obtain that
    \begin{equation}
        \int_\mathbb{B} |\langle f, \psi_{a,\theta,\tau}\rangle|^2 + |\langle f, \psi^*_{a,\theta,\tau}\rangle|^2 \frac{\mathrm{d}a \mathrm{d}\theta \mathrm{d}\tau}{a^3} = \int_{\mathbb{R}^2} |f(\varsigma)|^2 \mathrm{d}\varsigma,
    \end{equation}
    for any $f \in L^2(\mathbb{R}^2)$.
    Polarization then gives us that, for any $f, g \in L^2(\mathbb{R}^2)$,
    \begin{equation}
        \int_\mathbb{B} \langle f, \psi_{a,\theta,\tau}\rangle \langle g, \psi_{a,\theta,\tau} \rangle + \langle f, \psi^*_{a,\theta,\tau} \rangle \langle g, \psi^*_{a, \theta, \tau} \rangle \frac{\mathrm{d}a \mathrm{d}\theta \mathrm{d}\tau}{a^3} = \int_{\mathbb{R}^2} f(\varsigma) g(\varsigma) \mathrm{d}\varsigma.
    \end{equation}
    Now, by using the adjoint of the boostlet transform, we obtain
    \begin{equation}
        \int_{\mathbb{R}^2} \int_\mathbb{B} \langle f, \psi_{a,\theta,\tau}\rangle \psi_{a,\theta,\tau}(\varsigma) + \langle f, \psi^*_{a,\theta,\tau} \rangle \psi^*_{a,\theta,\tau}(\varsigma)  \frac{\mathrm{d}a \mathrm{d}\theta \mathrm{d}\tau}{a^3} g(\varsigma) \mathrm{d}\varsigma = \int_{\mathbb{R}^2} f(\varsigma) g(\varsigma) \mathrm{d}\varsigma.
    \end{equation}
    Since this holds for any $g \in L^2(\mathbb{R}^2)$, we have that the reproducing formula
    \begin{equation}
        f(\varsigma) = \int_\mathbb{B} \langle f, \psi_{a,\theta,\tau}\rangle \psi_{a,\theta,\tau}(\varsigma) + \langle f, \psi^*_{a,\theta,\tau} \rangle \psi^*_{a,\theta,\tau}(\varsigma) \frac{\mathrm{d}a \mathrm{d}\theta \mathrm{d}\tau}{a^3}
    \end{equation}
    holds in an $L^2$ sense for any $f \in L^2(\mathbb{R}^2)$.
    Consequently, $\psi$ is admissible.

\section{Proof of Theorem~\ref{thm:frame}}
\label{prf:thm:frame}
To prove this theorem, we follow~\cite[Ch. 3.3.2]{Daubechies1992} with some modifications to handle the sampling constraints induced by the boosts.

Let us consider the first part of the frame bound sum
\begin{align}
    &\sum_{\ell,n,m} |\langle f, \psi_{\ell,n,m}\rangle|^2 \\
    &\quad= \sum_{\ell,n,m} \left| \int_{\mathbb{R}^2} \widehat{f}(\xi) 2^{\ell} e^{-|n|\delta} \overline{\widehat{\psi}(2^\ell B_{n\delta} \xi)} e^{2\pi i 2^\ell e^{-|n|\delta} m^\transp \xi} \mathrm{d}\xi \right|^2 \\
    &\quad= 2^{2\ell} e^{-2|n|\delta} \sum_{\ell,n,m} \left| \int_{[0,2^{-\ell} e^{|n|\delta}]^2} e^{2\pi i 2^\ell e^{-|n|\delta} m^\transp \xi} \sum_s \widehat{f}(\xi + 2^{-\ell} e^{|n|\delta} s) \overline{\widehat{\psi}(2^\ell B_{n\delta} \xi + e^{|n|\delta} B_{n\delta} s)} \mathrm{d}\xi\right|^2,
\end{align}
where the $s$ ranges over all values of $\mathbb{Z}^2$ and we have used the fact that the complex exponential factor is periodic with a period of $2^{-\ell} e^{|n|\delta}$ in both directions.

We now consider the periodic function
\begin{equation}
    g(\xi) = \sum_s \widehat{f}(\xi + 2^{-\ell} e^{|n|\delta} s) \overline{\widehat{\psi}(2^\ell B_{n\delta} \xi + e^{|n|\delta} B_{n\delta} s)}
\end{equation}
defined on $[0, 2^{-\ell} e^{|n|\delta}]^2$ and note that the integral in the above sum is its Fourier series coefficient at $-m$.
We can thus use Parseval's identity to derive
\begin{equation}
    2^{2\ell} e^{-2|n|\delta} \sum_m \left| \int_{[0,2^{-\ell} e^{|n|\delta}]^2} e^{2\pi i 2^\ell e^{-|n|\delta} m^\transp \xi} g(\xi)\mathrm{d}\xi \right|^2 = \int_{[0, 2^{-\ell} e^{|n|\delta}]^2} |g(\xi)|^2 \mathrm{d}\xi.
\end{equation}
Plugging this into our expression for the square coefficient sum, we obtain
\begin{align}
    \sum_{\ell,n} \int_{[0,2^{-\ell}e^{|n|\delta}]^2} \left| \sum_s \widehat{f}(\xi + 2^{-\ell} e^{|n|\delta} s) \overline{\widehat{\psi}(2^\ell B_{n\delta} \xi + e^{|n|\delta} B_{n\delta} s)} \right|^2 \mathrm{d}\xi.
\end{align}
Expanding the square now gives
\begin{align}
    &\sum_{\ell,n} \int_{[0,2^{-\ell}e^{|n|\delta}]^2} \sum_{s,s'} \widehat{f}(\xi + 2^{-\ell} e^{|n|\delta} s) \overline{\widehat{\psi}(2^\ell B_{n\delta} \xi + e^{|n|\delta} B_{n\delta} s) \widehat{f}(\xi + 2^{-\ell} e^{|n|\delta} s')} \widehat{\psi}(2^\ell B_{n\delta} \xi + e^{|n|\delta} B_{n\delta} s') \mathrm{d}\xi \\
    &\quad=\sum_{\ell,n,s} \int_{\mathbb{R}^2} \widehat{f}(\xi) \overline{\widehat{\psi}(2^\ell B_{n\delta} \xi) \widehat{f}(\xi + 2^{-\ell} e^{|n|\delta} s)} \widehat{\psi}(2^\ell B_{n\delta} \xi + e^{|n|\delta} B_{n\delta} s) \mathrm{d}\xi,
\end{align}
where $s'$ again ranges over $\mathbb{Z}^2$ and we have extended the domain of the integral in the last step to include the sum over $s$ and renamed $s' \rightarrow s$.

Separating out the $s = (0, 0)^\transp$ terms, we obtain
\begin{align}
    &\int_{\mathbb{R}^2} |\widehat{f}(\xi)|^2 \mathrm{d}\xi \sum_{\ell,n} |\widehat{\psi}(2^{\ell} B_{n\delta} \xi)|^2  \\
    &\quad+\sum_{\ell,n} \sum_{s\neq(0,0)^\transp} \int_{\mathbb{R}^2} \widehat{f}(\xi)\overline{\widehat{f}(\xi + 2^{-\ell} e^{|n|\delta} s) \widehat{\psi}(2^\ell B_{n\delta} \xi)} \widehat{\psi}(2^\ell B_{n\delta} \xi + e^{|n|\delta} B_{n\delta} s) \mathrm{d}\xi.
\end{align}
We note that the eigenvalues of $B_{n\delta}$ are given by $e^{n\delta}$ and $e^{-n\delta}$, which means that $\|e^{|n|\delta} B_{n\delta} s\| \ge \|s\| \ge 1$.
Since we require that the mother boostlet $\psi$ has a Fourier support contained in a disk of diameter one, the product $\overline{\widehat{\psi}(2^\ell B_{n\delta} \xi)} \widehat{\psi}(2^\ell B_{n\delta} \xi + e^{|n|\delta} B_{n\delta} s)$ must always be zero whenever $s \neq (0, 0)^\transp$.
The second sum is therefore zero and we are left with the identity
\begin{align}
    \sum_{\ell,n,m} | \langle f, \psi_{\ell,n,m} \rangle |^2 = \int_{\mathbb{R}^2} |\widehat{f}(\xi)|^2 \mathrm{d}\xi \sum_{\ell,n} |\widehat{\psi}(2^\ell B_{n\delta} \xi)|^2.
\end{align}
The same result applies if we replace $\psi$ with $\psi^\ast$, and combining them gives
\begin{align}
    \sum_{\ell,n,m} | \langle f, \psi_{\ell,n,m} \rangle |^2 + | \langle f, \psi_{\ell,n,m}^\ast \rangle |^2 = \int_{\mathbb{R}^2} |\widehat{f}(\xi)|^2 \mathrm{d}\xi \sum_{\ell,n} |\widehat{\psi}(2^\ell B_{n\delta} \xi)|^2 + |\widehat{\psi}^\ast(2^\ell B_{n\delta} \xi)|^2.
\end{align}
Substituting in Eq.~\eqref{eq:frame-condition} then gives
\begin{align}
    A \|f\|^2 \le \sum_{\ell,n,m} | \langle f, \psi_{\ell,n,m} \rangle |^2 + | \langle f, \psi_{\ell,n,m}^\ast \rangle |^2  \le B \|f\|^2,
\end{align}
as desired.

\end{document}